\documentclass[twocolumn]{aastex61}
\submitjournal{ApJ}

\shorttitle{X-ray emissions of 3CRR quasars}
\shortauthors{Zhou \& Gu}

\begin{document}

\title{The composite X-ray spectrum of 3CRR Quasars}

\correspondingauthor{Minhua Zhou}
\email{zhoumh@shao.ac.cn}

\correspondingauthor{Minfeng Gu}
\email{gumf@shao.ac.cn}

\author{Minhua Zhou}
\affil{Key Laboratory for Research in Galaxies and Cosmology, Shanghai Astronomical Observatory, 
	Chinese Academy of Sciences, 80 Nandan Road, Shanghai 200030, China}
\affil{University of Chinese Academy of Sciences, 19A Yuquan Road, 
	Beijing 100049, China}

\author{Minfeng Gu}

\affiliation{Key Laboratory for Research in Galaxies and Cosmology, Shanghai Astronomical Observatory, 
	Chinese Academy of Sciences, 80 Nandan Road, Shanghai 200030, China}



\begin{abstract}

The reason for the difference in the composite X-ray spectrum between radio-loud quasars (RLQs) and radio-quiet quasars (RQQs) is still unclear. 
To study this difference, we built a new composite X-ray spectrum of RLQs by using Chandra X-ray data and Sloan Digital Sky Survey (SDSS) optical data for the sample of 3CRR quasars. 
We find the X-ray spectra of all 3CRR quasars except for 3C 351 have no soft X-ray excess and can be fitted with an absorbed power-law model well. 
Our composite X-ray spectrum is similar to that of \cite{ShangEtal2011apjs} for RLQs, showing higher hard X-ray and lower soft X-ray flux than the composite X-ray spectrum of RQQs. Most blazar-like 3CRR quasars have higher X-ray flux than the median composite X-ray spectrum, which could be related to the contribution of beamed jet emission at X-ray band. 
From the literature, we find that nineteen
3CRR quasars have extended X-ray emission related to radio jets, indicating inevitable contribution of jets at X-ray band. 
In contrast to RQQs, the X-ray photon index of 3CRR quasars does not correlate with the Eddington ratio. 
Our results suggest that the jet emission at X-ray band in RLQs could be related with the difference of composite X-ray spectrum between RLQs and RQQs.

\end{abstract}




\section{Introduction} \label{sec:intro}

X-ray emission appears to be nearly universal from active galactic nuclei (AGNs) and is often used to detect AGNs in various surveys \citep[see e.g.,][]{2005ARA&A..43..827B, 2012hcxa.confE.138W, 2013ApJ...773..125A, BrandtEtal2015A&ARv, 2017NewAR..79...59X}.
The intrinsic X-ray emission from AGNs usually originates in the immediate vicinity of the supermassive black hole, and it consists of several components including the primary X-ray emission with a high energy cut-off, soft X-ray excess, reflection and absorption components \cite[e.g.,][]{TurnerEtal2009aapr,MallickEtal2016MNRAS}. 
The primary X-ray emission is usually thought to be from corona by inverse Compton scattering of optical/UV photons \cite[e.g.,][]{HaardtEtal1993ApJ}. 
The origin of soft X-ray excess is still debated \citep{NodaEtal2013PASJ,MallickEtal2016MNRAS,JinEtal2017MNRAS,2018ApJ...863..178M}, including the model of thermal Comptonization in low temperature optically thick medium that separated from primary X-ray emission component \citep{MagdziarzEtal1998MNRAS,MarshallEtal2003AJ,DewanganEtal2007ApJ,DoneEtal2012MNRAS}, and the model of blurred reflection from ionized accretion disk \citep{FabianEtal2002MNRAS,CrummyEtal2006MNRAS,GarciaEtal2014ApJ}. 

To systematically study the emission in quasars, \citet[hereafter E94]{ElvisEtal1994apjs} produced the first all-band (from radio to X-ray bands) spectral energy distributions (SED) for a sample of quasars. 
Later, \citet[hereafter S11]{ShangEtal2011apjs} built the next generation atlas SED of quasars. 
These SEDs are composite spectra constructed for quasar samples based on multi-band observational data. With the median values within frequency bins (e.g., S11), the composite spectrum can be regarded as the representative of overall SED (e.g., the big blue bump, thermal IR bump, etc.) for the studied sample, thus the systematic comparison on the SED in different AGN populations can be readily performed by comparing their composite spectra.
Both of these two SEDs show prominent differences between radio-loud quasars (RLQs) and radio-quiet quasars (RQQs) at radio and X-ray bands, but have almost the same spectrum at other bands. While the difference in radio band could be most likely due to the presence of jets in RLQs, the reason for the difference in the X-ray band is still unclear. 
The possible reasons include the additional jet-related UV/X-ray flux \citep[e.g.,][]{WorrallEtal1987ApJ, MillerEtal2011ApJ}, the more ionized accretion disk \citep{2002MNRAS.332L..45B} or the X-ray emission from the hot Advection Dominated Accretion Flow (ADAF) within the truncated radius of accretion disk \citep{2014ARA&A..52..529Y} in radio-loud AGNs.

Many works found that radio-loud AGNs are more X-ray luminous and usually have harder X-ray spectra than radio-quiet AGNs \citep{ZamoraniEtal1981ApJ, WorrallEtal1987ApJ, 1987ApJ...323..243W,  2006ApJ...642..113G, 2011ApJ...740...29K, MillerEtal2011ApJ, 2013ApJ...763..109W, 2019MNRAS.482.2016Z}. 
As an early work, \cite{WorrallEtal1987ApJ} found that the relative X-ray brightness is greater for RLQs and suggested that an ``extra" X-ray emission would dominate the observed X-ray flux in the majority of RLQs with flat radio spectra. Comparing with RQQs \citep{SteffenEtal2006AJ}, \cite{MillerEtal2011ApJ} investigated the ``excess" X-ray luminosity of RLQs relative to RQQs as a function of radio loudness and luminosity. They proposed that the X-ray emission of RLQs may consist of disk/corona and jet-linked components.

Since RLQs usually have powerful jets, it could occur the jet emission will contribute to X-ray spectra \citep{WorrallEtal1987ApJ, MillerEtal2011ApJ}, especially in flat-spectrum radio quasars (FSRQs). As a matter of fact, \cite{GrandiEtal2004Sci} analyzed the X-ray spectrum of FSRQ 3C 273 with BeppoSAX data and found a model with jet plus Seyfert-like components can fit the spectrum very well. 
They found that the X-ray spectral index and soft excess flux have no correlation with total flux at X-ray band, but have a good correlation with the flux ratio of the jet to Seyfert components. It shows that the spectral index tends to be flatter when the jet-like component has more contribution to the total flux.

Since the jet is moving at small viewing angles in blazars (including FSRQs and BL Lac objects), the ``beaming effect'' is usually significant \citep{Antonucci1993ARAA, UrryEtal1995PASP}. 
The jet emission in blazars can be significantly boosted due to the ``beaming effect''. In other words, the observed emission can be much larger than the intrinsic one caused by the Doppler effect from relativistic jet speed.
This results in a nontrivial jet contribution at all bands, like X-ray band studied in this work, as indicated in 3C 273 above. 
Although the blazars were claimed to be excluded in both E94 and S11 work, we found 22 blazars out of 58 RLQs in S11 sample, and 7 blazars out of 18 RLQs in E94 by checking with BZCAT catalog \citep{MassaroEtal2009aap, 2015Ap&SS.357...75M} Edition 5.0.0\footnote{\url{http://www.asdc.asi.it/bzcat/}}. 
In addition, the previous works on the X-ray difference between RLQs and RQQs may also include blazars in their RLQs sample \cite[e.g.,][]{WorrallEtal1987ApJ, MillerEtal2011ApJ}.
In this work, we intend to revisit the difference of composite X-ray spectrum between RLQs and RQQs, by building a new composite X-ray spectrum for a sample of non-blazar RLQs. 
The ideal way to revisit the composite X-ray spectrum of RLQs is to use the pure thermal emission by decomposing from jet emission in X-ray spectra. However, this is usually hard to achieve for a sample study. Alternatively, we can minimize the contribution of jet emission by carefully selecting the RLQs sample.

In Section \ref{sec:sample}, we introduce our sample. 
The observational data of our sample, and data reduction on multi-band especially for Chandra or XMM-Newton observations, are shown in Section \ref{secdata}. 
Our results of the composite spectrum of 3CRR quasars are given in Section \ref{sec:result}, and discussed in Section \ref{discussion}. 
Finally, our results are summarized in Section \ref{summary}. 
Throughout this paper, the cosmological
parameters $H_{0}=70\rm ~km~ s^{-1}~Mpc^{-1}$, $\Omega_{\rm m}=0.3$ and $\Omega_{\Lambda}=0.7$ are adopted.
Photon index $\Gamma$ is defined as
$A(E) = K E^{-\Gamma}$, where $K$ is photons at $1~\rm keV$ and E is photon energy. The spectral index $\alpha$ is defined as $f_{\rm \nu} \varpropto \nu^{-\alpha}$ with $f_{\rm \nu}$ being the flux density at frequency $\nu$.

\section{Sample} \label{sec:sample}

3CRR catalog lists the brightest radio sources in northern sky, and it was selected at low radio frequency 178 MHz with flux density brighter than $10.9~\rm Jy$ \citep{LaingEtal1983mnras}. Due to the low-frequency selection, 3CRR sample is dominated by steep-spectrum sources, therefore their SEDs are less likely dominated by beamed jets. 3CRR catalog consists of 43 quasars and 130 radio galaxies.

As one of the few samples with almost complete multi-waveband observations, the sample of 3CRR quasars is ideal for us to investigate the overall SED, especially the X-ray spectra in the present work. 
The extensive studies have been presented in various works, such as, at radio \cite[e.g.,][]{BedfordEtal1981mnras, BridleEtal1994aj, Fernini2007aj, Fernini2014apjs}, submillimeter-wave \cite[e.g.,][]{HaasEtal2004aap}, infrared \cite[e.g.,][]{ HaasEtal2008apj,Gurkan2014mnras}, optical \cite[e.g.,][]{LehnertEtal1999apjs,AarsEtal2005aj} and X-ray bands \cite[e.g.,][]{TananbaumEtal1983apj,WorrallEtal2004MNRAS, HardcastleEtal2009MNRAS, WilkesEtal2013ApJ, MassaroEtal2013apjs}.

Our sample of 43 3CRR quasars is shown in Table \ref{table:sample}. Most of these quasars are lobe dominate quasars (LDQs). The sample includes two well know FSRQs (3C 345 and 3C 454.3) \citep{Healey2007ApJS}, and four blazars of uncertain type (3C 207, 3C 216, 3C 309.1 and 3C 380) \citep{MassaroEtal2009aap, 2015Ap&SS.357...75M}.  
Our sources cover a broad range of redshift, from 0.311 to 2.012. The flux density of the radio core from Very Large Array (VLA) observations at 5 GHz ranges from 1.1 mJy to 12.2 Jy. 
Calculated using VLA 5 GHz core flux, the radio loudness $\log R$ varies from 0.5 to 4.3 \citep[$R=f_{5~\rm GHz}/f_{4400~\rm \AA}$,][]{1989AJ.....98.1195K}.
The Chandra X-ray data is available for all quasars \citep{2018ApJS..234....7M, 2018ApJS..235...32S}. The majority of 3CRR quasars were detected in SDSS \citep{2009ApJS..182..543A, 2015ApJS..219...12A} and Two Micron All Sky Survey \citep[2MASS,][]{2003tmc..book.....C} photometric catalogs.

\startlongtable
\begin{deluxetable*}{l|lcrcccllc}
	\tablecaption{The sample of 3CRR Quasars \label{table:sample}}
	\tablewidth{0pt}
	\tablehead{
		\colhead{Name} & \colhead{IAU name} & \colhead{$ z $} & \colhead{Core flux} & \colhead{$\log R$} & \colhead{Radio class} & \colhead{Ref.} & \colhead{CXO ID} & \colhead{SDSS} & \colhead{2MASS} 
	}
	\decimalcolnumbers
	\startdata
    3C 9    &  0017+154  & 2.012 & 4.9   & 1.2  & LDQ   & A05  & 1595  & 2000/11/30 & y \\
	3C 14   &  0033+183  & 1.469 & 10.6  & 2.0  & LDQ   & A05   & 9242  & 2008/11/02 & y \\
	3C 43   &  0127+233  & 1.470 & $<61.0$ & 3.3  & CDQ & L83     & 9324  & 2004/09/21 & n \\
	3C 47   &  0133+207  & 0.425 & 73.6  & 2.4  & LDQ   & A05   & 2129  & 2009/01/26 & y \\
	3C 48   &  0134+329  & 0.367 & 896.0   & 2.9  & CDQ & L83     & 3097  & 2008/10/31 & y \\
	3C 68.1  &  0229+341  & 1.238 & 1.1  & 0.9  & LDQ   & A05   & 9244  & NED   & y \\
	3C 138  &  0518+165  & 0.759 & 94.0    & 2.8  & CDQ & L83     & 14996 & 2006/11/01 & y \\
	3C 147  &  0538+498  & 0.545 & 2500.0  & 3.2  & CDQ & L83     & 14997 & SSDC  & y \\
	3C 175  &  0710+118  & 0.768 & 23.5  & 1.5  & LDQ   & A05   & 14999 & NED   & y \\
	3C 181  &  0725+147  & 1.382 & 6.0     & 1.5  & LDQ & A05     & 9246  & 2006/11/23 & y \\
	3C 186  &  0740+380  & 1.063 & 15.0    & 1.9  & CDQ & L83     & 9774  & 2000/04/04 & y \\
	3C 190  &  0758+143  & 1.197 & 73.0    & 2.9  & LDQ & A05     & 17107  & 2004/12/15 & y \\
	3C 191  &  0802+103  & 1.952 & 42.0    & 2.5  & LDQ & A05     & 5626  & 2005/03/10 & y \\
	3C 196  &  0809+483  & 0.871 & 7.0     & 1.5  & LDQ & A93     & 15001 & 2000/04/25 & y \\
	3C 204  &  0833+654  & 1.112 & 26.9  & 2.0  & LDQ   & A05   & 9248  & 2004/10/15 & y \\
	3C 205  &  0835+580  & 1.534 & 20.0    & 1.5  & LDQ & A05     & 9249  & 2003/10/24 & y \\
	3C 207  &  0838+133  & 0.684 & 510.0   & 3.5  & LDQ & A05 & 2130  & 2005/03/10 & y \\
	3C 208  &  0850+140  & 1.109 & 51.0    & 2.3  & LDQ & A05     & 9250  & 2005/11/10 & y \\
	3C 212  &  0855+143  & 1.049 & 150.0   & 3.2  & LDQ & A05     & 434   & 2005/12/06 & y \\
	3C 215  &  0903+169  & 0.411 & 16.4  & 2.0  & LDQ   & A05   & 3054  & 2005/01/19 & y \\
	3C 216  &  0906+430  & 0.668 & 1050.0  & 4.1  & CDQ & L83 & 15002 & 2011/11/21 & y \\
	3C 245  &  1040+123  & 1.029 & 910.0   & 3.6  & LDQ & A05     & 2136  & 2003/03/31 & y \\
	3C 249.1 &  1100+772  & 0.311 & 71.0   & 1.7  & LDQ & A05     & 3986  & 2006/04/21 & y \\
	3C 254  &  1111+408  & 0.734 & 19.0    & 1.8  & LDQ & A93     & 2209  & 2003/04/01 & y \\
	3C 263  &  1137+660  & 0.652 & 157.0   & 2.1  & LDQ & A05     & 2126  & 2001/03/19 & y \\
	3C 268.4  &  1206+439  & 1.402 & 50.0  & 2.2  & LDQ & A05     & 9325  & 2003/04/25 & y \\
	3C 270.1  &  1218+339  & 1.519 & 190.0 & 3.1  & LDQ & A05     & 13906 & 2004/04/25 & y \\
	3C 275.1  &  1241+166  & 0.557 & 130.0 & 3.0  & LDQ & A05     & 2096  & 2005/06/06 & y \\
	3C 287  &  1328+254  & 1.055 & 2998.0  & 4.3  & CDQ & L83     & 3103  & 2004/12/21 & y \\
	3C 286  &  1328+307  & 0.849 & 5554.0  & 4.3  & CDQ & L83     & 15006 & 2004/05/12 & y \\
	3C 309.1  &  1458+718  & 0.904 & 2350.0& 3.7  & CDQ & L83 & 3105  & SSDC  & y \\
	3C 325  &  1549+628  & 0.860 & 2.4   & 1.8  & LDQ   & This work & 4818  & 2004/06/15 & n \\
	3C 334  &  1618+177  & 0.555 & 111.0   & 2.4  & LDQ & A05     & 2097  & 2004/04/22 & y \\
	3C 336  &  1622+238  & 0.927 & 20.4  & 2.2  & LDQ   & A05   & 15008 & 2004/05/13 & y \\
	3C 343  &  1634+628  & 0.988 & $<300.0$& 3.9  & CDQ & L83     & 15011 & 2006/05/01 & n \\
	3C 345  &  1641+399  & 0.594 & 8610.0  & 3.7  &  CDQ& L83  & 2143  & 2001/05/23 & y \\
	3C 351  &  1704+608  & 0.371 & 6.5   & 0.5  & LDQ   & A05   & 435   & 2000/04/04 & y \\
	4C 16.49  &  1732+160  & 1.296 & 16.0  & 2.0  & LDQ & A05     & 9262  & NED   & n \\
	3C 380  &  1828+487  & 0.691 & 7447.0  & 4.1  & CDQ & L83 & 3124  & SSDC  & y \\
	3C 432  &  2120+168  & 1.785 & 7.5   & 1.6  & LDQ   & A05   & 5624  & 2008/09/24 & y \\
	3C 454  &  2249+185  & 1.757 & $<200.0$& 3.2  & CDQ & L83     & 21403 & 2009/01/21 & n \\
	3C 454.3  &  2251+158  & 0.859 & 12200.0 & 3.6 & CDQ& L83  & 4843  & 2008/10/02 & y \\
	3C 455  &  2252+129  & 0.543 & 1.4   & 1.5  & LDQ   & F03   & 15014 & 2001/09/18 & y \\
	\enddata
	\tablecomments{
		Column (1): 3CRR name; Column (2): IAU name; Column (3): Redshift \citep{LaingEtal1983mnras}; Column (4): VLA $5~\rm GHz$ core flux in $\rm mJy$ with all from \cite{LaingEtal1983mnras}, except for 3C 287 \citep{Laurent-MuehleisenEtal1997AAS}; Column (5): Radio loudness, $R=f_{5\rm GHz}/f_{4400\rm \AA}$, where $f_{5\rm GHz}$ is rest-frame VLA $5~\rm GHz$ core flux; Column (6): Radio classification, LDQ for Lobe dominated quasars, CDQ for Core dominated quasars; 
		Column (7): References for radio classification, A05 - \cite{AarsEtal2005aj}, A93 - \cite{1993AJ....105.2054A}, F03 - \cite{FanEtal2003}, L83 - \cite{LaingEtal1983mnras}, the classification of 3C 325 is based on the core dominance value of 0.003 calculated from VLA measurements in \cite{1997AJ....114.2292F};
		Column (8): Chandra observation ID; Column (9): The observational time of SDSS photometric data. When SDSS data is unavailable, the optical data were collected from NED or SSDC; Column (10): Available 2MASS data, y for $yes$, n for $no$.}
\end{deluxetable*}

\section{Data and reduction} \label{secdata}

To create the composite X-ray spectrum of 3CRR quasars, we collected available optical/IR and X-ray data for all our sources.

\subsection{Optical and Infrared data}

There are 37 quasars detected by Sloan Digital Sky Survey (SDSS) at $ugriz$ bands. For the remaining six sources (3C 68.1, 3C 147, 3C 175, 3C 309.1, 4C 16.49 and 3C 380), the photometric data at optical and/or near infrared bands were collected from NASA/IPAC Extragalactic Database (NED\footnote{\url{http://ned.ipac.caltech.edu/}}) or Space Science Data Center (SSDC\footnote{\url{http://www.asdc.asi.it/}}) (see Table \ref{table:sample}). 

We obtained SDSS photometric data from SDSS DR12\footnote{\url{http://skyserver.sdss.org/dr12/en/home.aspx}}, including fiber magnitude, fiber magnitude error and extinction values at all bands. After correcting the Galactic extinction, we converted SDSS magnitudes $m_{\rm AB}$ to flux densities $f_{\rm \nu}$ using the zeropoint flux density of  $f_{\rm \nu 0}$=3631 $\rm Jy$ \citep{OkeEtal1983ApJ}.

To produce the composite X-ray spectrum of 3CRR quasars, we followed the same method with S11, in which the rest-frame $4215~\rm \AA$ was used as a reference
for multi-band flux densities. 
In this case, we required the available photometric data to cover the rest-frame $ 4215~\rm \AA$. 
In high-redshift quasars with $ 4215\rm \AA$ not covered by SDSS or NED/SSDC optical data, the near infrared data were added (see details in Section \ref{sec:3CRR_SED}). 
We used 2MASS data archived by Infrared Science Archive (\href{https://irsa.ipac.caltech.edu/frontpage/}{IRSA}) for all quasars, except for 3C 270.1, in which the 2MASS data was taken from \citet{KrawczykEtal2013apjs}.
The 2MASS magnitudes were converted to flux densities with zeropoint flux densities 1594, 1024, and 666.7 Jy for J, H, and Ks bands, respectively \citep{CohenEtal2003AJ}.

The near- to mid-IR data at 3.4, 4.6, 12 and 22$\rm \mu m$ are available for all 3CRR quasars from the WISE all-sky data release \citep{2010AJ....140.1868W}. In principle, these data can be used to build broadband optical to IR SEDs for sample sources. However, we mainly focus on the flux density at 4215$\rm \AA$ and the optical SED, thus the WISE data were not used in this work.

\subsection{X-ray data} \label{subsec:xraydata}

3CRR quasars have been observed by various X-ray telescopes, such as ROSAT, NuSTAR, XMM-Newton, and Chandra. In this work, we prefer to use Chandra data for several reasons. 
First of all, all 3CRR quasars have been observed by Chandra telescope. We  can reduce the data in a uniform way with the same calibration so not to take into account corrections due to intercalibrations of different satellites. 
Secondly, Chandra telescope has higher angular resolution than XMM-Newton. This high angular resolution ($\sim$ 0.5\arcsec) has an incomparable advantage to isolate the core emission from the elongated jets. 
Thirdly, Chandra covers from 0.3 to 10 $\rm keV$ band. It thus can be used to study the difference at both soft and hard X-ray bands between RLQs and RQQs. 
For the sources with multiple observations from Chandra or SDSS, we chose those X-ray and optical data with the closest time separation. However, when all the time separation longer than one year, we selected X-ray data with longer exposure time. The selected X-ray and optical data are shown in Table \ref{table:sample}. 

The Chandra X-ray data of most 3CRR quasars have already been analyzed in the literature \cite[e.g.,][]{CrawfordEtal2003MNRAS,HardcastleEtal2004ApJ,CrostonEtal2005ApJ,MassaroEtal2010ApJ,MassaroEtal2012ApJS,MassaroEtal2013apjs,MassaroEtal2015ApJS}. 
However, these works mostly focused on either the extended X-ray emission \citep[e.g.,][]{HardcastleEtal2004ApJ} or only the X-ray flux \citep[e.g.,][]{MassaroEtal2015ApJS}.
The X-ray spectra of many 3CRR quasars have been studied with detailed spectral fitting in various works \citep{BrunettiEtal2002aap,AldcroftEtal2003ApJ,DonahueEtal2003ApJ,FabianEtal2003MNRAS,BelsoleEtal2006MNRAS,HardcastleEtal2006MNRAS,SiemiginowskaEtal2008ApJ,HardcastleEtal2009MNRAS,SiemiginowskaEtal2010ApJ,WilkesEtal2012ApJ,WilkesEtal2013ApJ}. However, in these works, the spectra were analyzed by different groups and extracted from different regions likely causing systematic difference between individual objects. Instead of directly taking the results from the literature, we decided to re-analyze the selected Chandra data for 3CRR quasars in a uniform way.

All 3CRR quasars were observed by Chandra X-ray Observatory (CXO) Advanced CCD Imaging Spectrometer (ACIS). We used Chandra Interactive Analysis of Observations software (CIAO) v4.8 and Chandra Calibration Database (CALDB) version 4.7.1 to reduce the data step by step with CIAO threads\footnote{\url{http://cxc.harvard.edu/ciao/threads/index.html}}. We used $chandra\_repro$ script to reprocess data and to create a new level 2 event file and a new bad pixel file. After that, we checked all sources for background flares and filtered energy between $0.3-10~\rm keV$. In the end, we extracted spectrum and response files from a source-centered circle of radius $2.5\arcsec$ with $specextract$ script. For the background region, we used $20-30\arcsec$ annulus if there is no other sources in this region, otherwise, we used an arbitrary background region around the quasar.

\begin{figure}[htb!]
	\includegraphics[width=1.0\columnwidth]{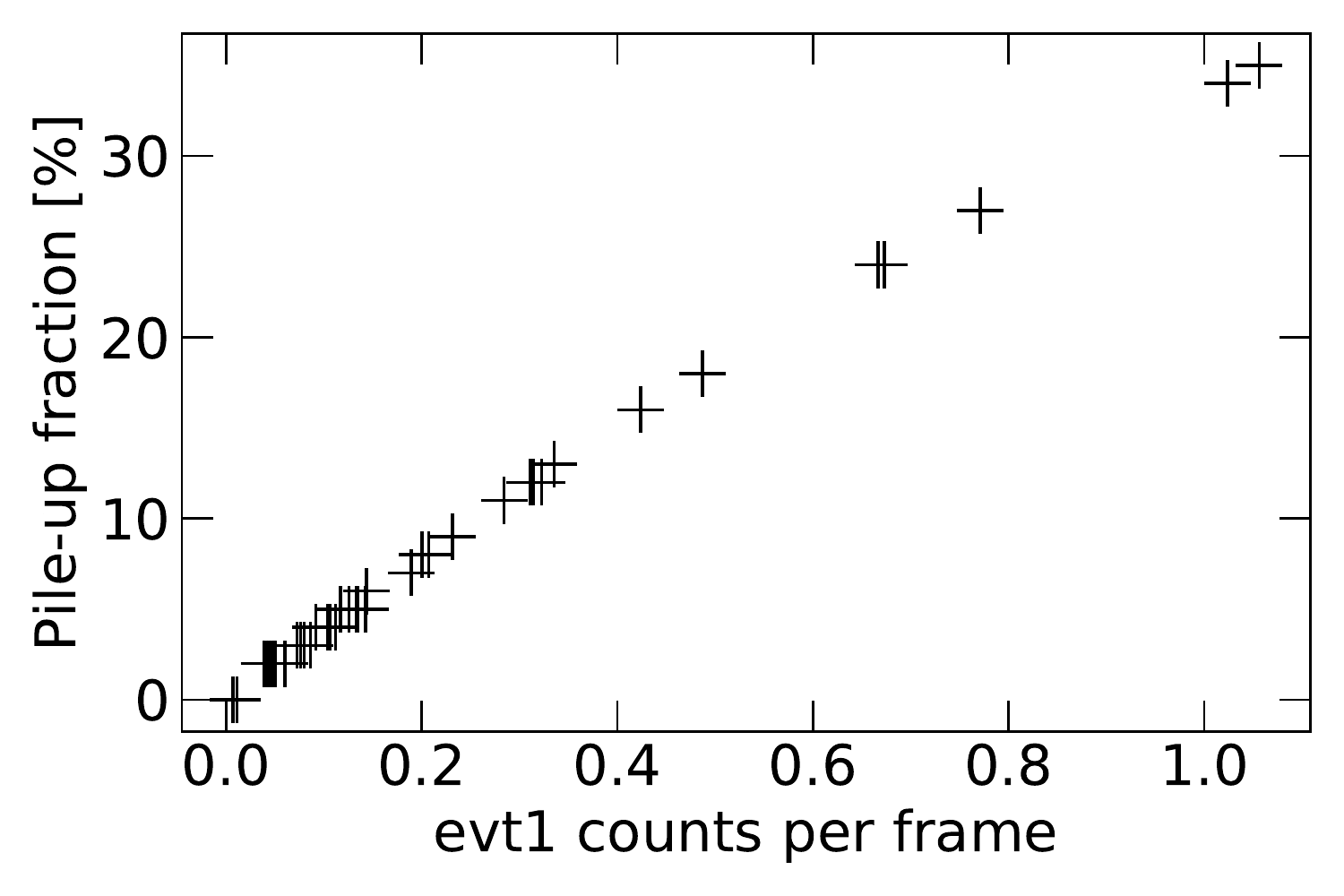}
	\caption{Pile-up fraction versus evt1 counts per frame.\label{fig:pileup}}
\end{figure}

Before we analyzed the spectra, we carefully studied the pile-up effect, i.e., when two or more photons are detected as a single event. This effect will cause a distortion of the energy spectrum, such as two low energy photons pile to a high energy photon. To confirm which source we should use $pileup$ model in Sherpa, we used PIMMS\footnote{\url{http://cxc.harvard.edu/toolkit/pimms.jsp}} to estimate the degree of pile-up. After inputting a series of parameters including the count rate of evt1 file in $1.5\arcsec$ circle centered on source calculated by $Funtools$, a power-law spectrum with photon index $\Gamma = 2$, redshift, Galactic {H\,{\footnotesize I}} Column Density (nH) \citep{DickeyEtal1990araa} and frame time of each observation, PIMMS returns the pile-up fraction. As did in \cite{MassaroEtal2012ApJS}, we plotted the relation of pile-up fraction with evt1 counts per frame in Figure \ref{fig:pileup}. A strong correlation was found between pile-up fraction and evt1 counts per frame in $1.5\arcsec$ circle. 
In this paper, we consider the pile-up effect according to the last version of Chandra Cycle 22 Proposers' Observatory Guides\footnote{\url{https://cxc.harvard.edu/proposer/POG/}}.

In the case that the pile-up effect can't be ignored, there are two ways to extract and analyze the X-ray spectrum. One is to extract the spectrum from an annular area by excluding the piled-up core, for example, $0.5-2.0\arcsec$ \cite[e.g.,][]{GambillEtal2003aap,WorrallEtal2004MNRAS}. 
The other is to fit the spectra extracted from the central region, which includes piled-up core, with $jdppileup$ model \citep{DavisEtal2001apj} in Sherpa. In this work, the latter method was adopted for those spectra extracted from $2.5\arcsec$ circle when the pile-up effect is significant, i.e., evt1 count rate higher than $0.2~\rm counts/frame$.

We used Sherpa \citep{FreemanEtal2001SPIE} to fit all X-ray spectra for 3CRR quasars, and two statistical methods were used (see Table \ref{table:3cfit1}). One is $cstat$ for the case that the photon count from source-centered $2.5 \arcsec$ circle is less than 200, and thus the X-ray spectrum was basically unbinned. The other is $chi2xspecvar$ for the sources with enough exposure time, and the X-ray spectra were binned with 15 counts per bin. 

\section{Results} \label{sec:result}

\subsection{X-ray spectra} \label{subsec:Xresults}

\begin{figure}[htb!]
	\includegraphics[width=0.5\textwidth]{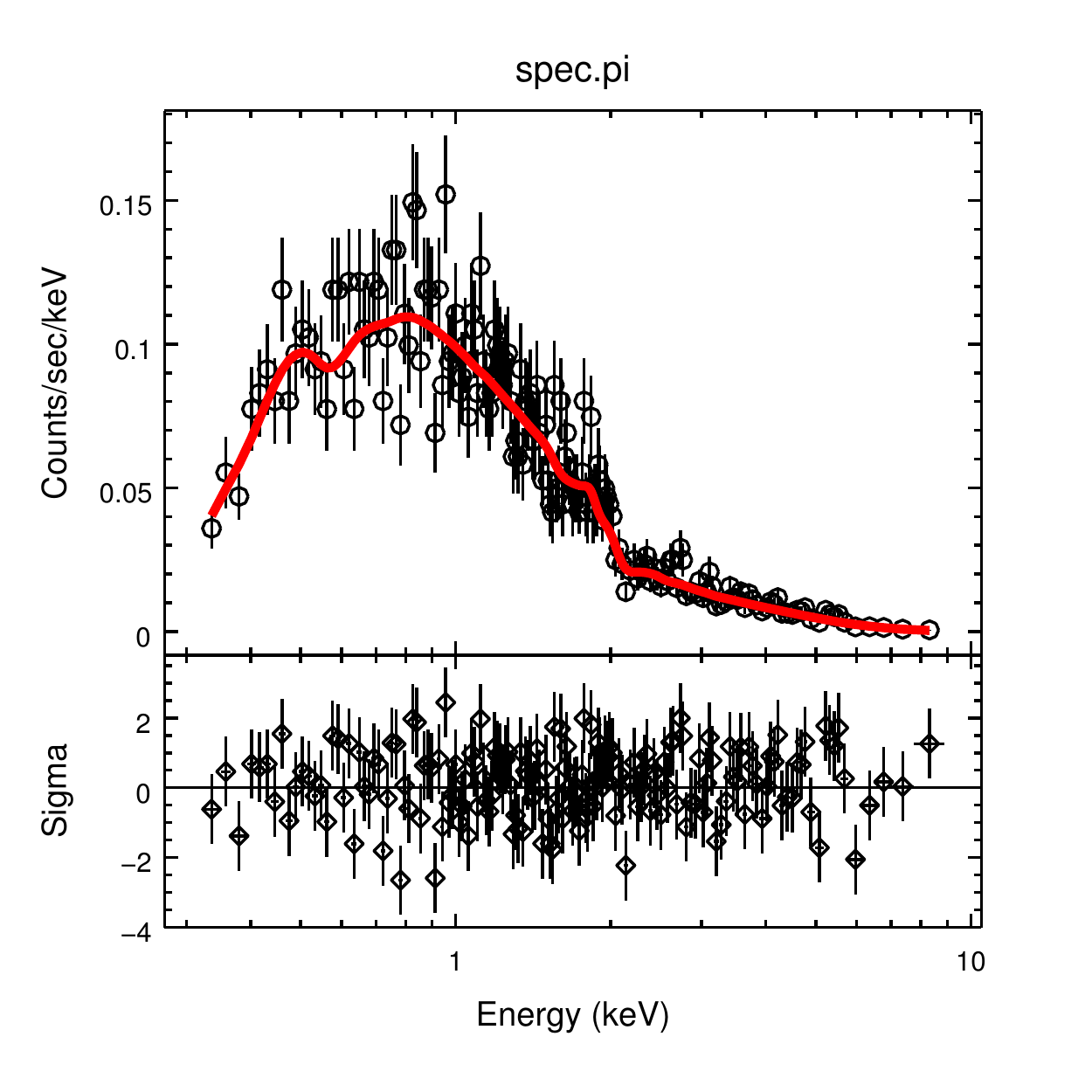}
	\caption{An example of X-ray spectral fitting. A single power-law can fit the spectrum well in 3C 275.1.\label{fig:3c47}}
\end{figure}

The Chandra X-ray spectra extracted from source-centered $2.5\arcsec$ circle were fitted with three models, including intrinsic absorbed power-law model with fixed Galactic absorption ($phabs*zphabs*powerlaw$), absorbed broken power-law model ($phabs*zphabs*bknpower$) and absorbed double power-law model ($phabs*zphabs*(powerlaw1+powerlaw2)$). 
We found all spectra except for 3C 351 can be best fitted by a single power-law model. As an example, the X-ray spectral fitting for 3C 275.1 is shown in Figure \ref{fig:3c47}. The X-ray spectrum of 3C 351 shows a prominent soft X-ray excess when fitted with the absorbed power-law model, and can be well fitted with a warm absorbed power-law model (see details in Appendix \ref{subsec:3c351fit}). The X-ray photon index is obtained from the spectral fit for all quasars except for 3C 68.1, in which it was fixed to $\Gamma=1.9$ due to severe absorption (see Appendix \ref{subsubsec:3c68}). 
The results of X-ray spectral fitting are shown in Table \ref{table:3cfit1}, including X-ray photon index, neutral Hydrogen column density, the absorption-corrected flux at $0.3-2.0~\rm keV$, $2.0-10.0~\rm keV$ and $0.3-10.0~\rm keV$, etc.. 
The measurement errors of model parameters (e.g., $ \Gamma $ and power-law normalization) were estimated with $ conf $ script in Sherpa. The flux and its uncertainties were calculated with $ sample\_flux $ script in Sherpa.

\subsubsection{Compare with other works}

There are various works dedicated to detailed spectral analysis on the Chandra data for some of 3CRR quasars with also simple power-law model. \cite{WilkesEtal2013ApJ} extracted X-ray photons from source-centered 2.2\arcsec circle and presented the results of X-ray spectral fitting for 
nine 3CRR quasars (3C 9, 3C 186, 3C 191, 3C 205, 3C 212, 3C 245, 3C 270.1, 3C 287, and 3C 432). We found that the X-ray photon index of their work is consistent with our results within errors. 
The spectral analysis on three 3CRR quasars (3C 47, 3C 215, 3C 249.1) have been performed by \cite{HardcastleEtal2006MNRAS}. With extracted X-ray spectra from 1.25\arcsec circle, the results of their work are generally in good agreement with our results.
 
\cite{BelsoleEtal2006MNRAS} and \cite{GambillEtal2003aap} studied the X-ray properties for nine quasars (3C 207, 3C 254, 3C 263, 3C 275.1, 3C 309.1, 3C 334, 3C 345, 3C 380, and 3C 454.3). From their results, the fitted X-ray spectra are in general flatter than our results. The difference is likely due to the PSF effect of Chandra/ACIS detector that the harder X-ray has broader PSF, since the central piled-up region was excluded to avoid pile-up effect in their studies. 
This possibility is supported by our simulation, in which we found flatter X-ray spectra in outer region than inner region (see Appendix \ref{sec:in-out}). The K$ - $S test shows significantly different distributions of X-ray photon index in two regions.
Using the same method to avoid the pile-up effect as \cite{BelsoleEtal2006MNRAS}, \cite{HardcastleEtal2009MNRAS} presented the slightly flatter photon index than our results for 3C 48, however the result of 3C 325 is same as ours, in which there is no pile-up effect. 
On the other hand, the similar X-ray photon index with our results have also been found in the sources with pile-up effect, including 3C 48 in \cite{SiemiginowskaEtal2008ApJ}, 3C 254 in \cite{DonahueEtal2003ApJ}, 3C 263 in \cite{HardcastleEtal2002ApJ}, and 3C 454.3 in \cite{2007ApJ...662..900T}. 

\begin{longrotatetable}
	\begin{deluxetable*}{lrrr|rrrlrll|rrr}
		\tablecaption{The results of X-ray spectral fitting \label{table:3cfit1}}
		\tablewidth{700pt}
		\tabletypesize{\scriptsize}
		
		\tablehead{
			\multicolumn{4}{c}{ } & \multicolumn{7}{c}{$ 2.5\arcsec $ circle} & \multicolumn{3}{c}{$log_{10}~ f ~(\rm erg~ cm^{-2}~s^{-1})$} \\
			\colhead{Name} & \colhead{$ z $} &
			\colhead{nH} & \colhead{CXO ID}  &
			\colhead{$ z $.nH} & \colhead{jdp.a} & 
			\colhead{jdp.f} & \colhead{$\Gamma$} & 
			\colhead{Norm.} & \colhead{$\chi^2$/dof.} & \colhead{Stat.} & 
			\colhead{$0.3-2.0~\rm keV$} & \colhead{$2.0-10~\rm keV$} & \colhead{$0.3-10.0~ \rm keV$} 
		} 
		\decimalcolnumbers
		\startdata
    	3C 9           & 2.012         & 3.57          & 1595          & $ 1.45E-02 $  &               &               & $1.61_{-0.08}^{+0.12} $ & $ 0.44_{-0.02}^{+0.04} $  & 1.48/27       & chi2          & $-12.90^{+0.07}_{-0.06} $ & $-12.69^{+0.16}_{-0.12} $ & $-12.48^{+0.11}_{-0.09} $ \\
		3C 14          & 1.469         & 4.04          & 9242          & $ 2.54E-01 $  &               &               & $1.53_{-0.18}^{+0.26} $ & $ 1.12_{-0.17}^{+0.31} $  & 0.71/11       & chi2          & $-12.51^{+0.10}_{-0.09} $ & $-12.24^{+0.23}_{-0.15} $ & $-12.05^{+0.16}_{-0.12} $ \\
		3C 43          & 1.470         & 6.65          & 9324          & $ 1.04E-03 $  &               &               & $1.53_{-0.15}^{+0.25} $ & $ 0.80_{-0.08}^{+0.20} $  & 0.78/114      & cstat         & $-12.65^{+0.04}_{-0.04} $ & $-12.37^{+0.12}_{-0.09} $ & $-12.19^{+0.08}_{-0.06} $ \\
		3C 47          & 0.425         & 4.95          & 2129          & $ 4.15E-02 $  & 0.65          & 0.94          & $1.87_{-0.22}^{+0.21} $ & $ 9.39_{-3.52}^{+2.66} $  & 1.08/288      & chi2          & $-11.55^{+0.14}_{-0.14} $ & $-11.54^{+0.25}_{-0.17} $ & $-11.24^{+0.16}_{-0.14} $ \\
		3C 48          & 0.367         & 4.34          & 3097          & $ 0.00E+00 $  & 0.86          & 0.85          & $2.32_{-0.01}^{+0.16} $ & $ 10.65_{-0.20}^{+1.66} $  & 1.24/171      & chi2          & $-11.45^{+0.01}_{-0.01} $ & $-11.76^{+0.02}_{-0.02} $ & $-11.28^{+0.01}_{-0.01} $ \\
		3C 68.1        & 1.238         & 5.38          & 9244          & $ 5.31E+00 $  &               &               & $1.90$       & $ 0.55_{-0.12}^{+0.15} $  & 0.96/39       & cstat         & $-12.78^{+0.10}_{-0.10} $ & $-12.78^{+0.10}_{-0.10} $ & $-12.49^{+0.11}_{-0.10} $ \\
		3C 138         & 0.759         & 21.90         & 14996         & $ 2.56E-03 $  & 0.80          & 0.88          & $1.46_{-0.11}^{+0.24} $ & $ 3.83_{-0.42}^{+0.99} $  & 1.80/18       & chi2          & $-11.97^{+0.06}_{-0.06} $ & $-11.63^{+0.10}_{-0.10} $ & $-11.48^{+0.08}_{-0.07} $ \\
		3C 147         & 0.545         & 20.40         & 14997         & $ 2.02E-01 $  &               &               & $1.85_{-0.22}^{+0.24} $ & $ 2.03_{-0.42}^{+0.55} $  & 0.75/106      & cstat         & $-12.22^{+0.11}_{-0.11} $ & $-12.19^{+0.24}_{-0.15} $ & $-11.90^{+0.14}_{-0.11} $ \\
		3C 175         & 0.768         & 10.30         & 14999         & $ 2.45E-03 $  & 0.02          & 0.85          & $1.52_{-0.07}^{+0.40} $ & $ 3.42_{-0.25}^{+0.81} $  & 0.96/17       & chi2          & $-12.02^{+0.03}_{-0.03} $ & $-11.73^{+0.08}_{-0.07} $ & $-11.55^{+0.06}_{-0.05} $ \\
		3C 181         & 1.382         & 6.63          & 9246          & $ 2.58E-04 $  &               &               & $1.72_{-0.10}^{+0.17} $ & $ 0.97_{-0.07}^{+0.13} $  & 1.14/117      & cstat         & $-12.56^{+0.04}_{-0.04} $ & $-12.41^{+0.10}_{-0.08} $ & $-12.18^{+0.06}_{-0.05} $ \\
		3C 186         & 1.063         & 5.11          & 9774          & $ 1.12E-03 $  &               &               & $1.88_{-0.03}^{+0.05} $ & $ 0.97_{-0.04}^{+0.01} $  & 1.12/172      & chi2          & $-12.54^{+0.02}_{-0.02} $ & $-12.52^{+0.03}_{-0.03} $ & $-12.23^{+0.02}_{-0.02} $ \\
		3C 190         & 1.197         & 3.04          & 17107         & $ 1.98E-02 $  &               &               & $1.61_{-0.05}^{+0.06} $ & $ 0.64_{-0.03}^{+0.04} $  & 1.12/85       & chi2          & $-12.74^{+0.02}_{-0.02} $ & $-12.52^{+0.04}_{-0.04} $ & $-12.32^{+0.03}_{-0.03} $ \\
		3C 191         & 1.952         & 2.45          & 5626          & $ 1.53E-01 $  &               &               & $1.68_{-0.10}^{+0.11} $ & $ 0.54_{-0.04}^{+0.05} $  & 0.88/38       & chi2          & $-12.81^{+0.04}_{-0.04} $ & $-12.64^{+0.09}_{-0.07} $ & $-12.42^{+0.06}_{-0.05} $ \\
		3C 196         & 0.871         & 4.82          & 15001         & $ 2.36E+00 $  &               &               & $1.67_{-0.38}^{+0.40} $ & $ 1.43_{-0.52}^{+0.84} $  & 0.84/71       & cstat         & $-12.38^{+0.20}_{-0.19} $ & $-12.24^{+0.49}_{-0.25} $ & $-11.99^{+0.33}_{-0.20} $ \\
		3C 204         & 1.112         & 4.59          & 9248          & $ 3.18E-01 $  & 1.00          & 1.00          & $2.15_{-0.22}^{+0.14} $ & $ 2.41_{-0.32}^{+0.29} $  & 0.96/156      & cstat         & $-12.11^{+0.12}_{-0.13} $ & $-12.29^{+0.16}_{-0.14} $ & $-11.89^{+0.13}_{-0.15} $ \\
		3C 205         & 1.534         & 4.48          & 9249          & $ 4.29E-01 $  & 1.00          & 1.00          & $1.83_{-0.17}^{+0.06} $ & $ 1.78_{-0.21}^{+0.12} $  & 0.94/52       & chi2          & $-12.28^{+0.05}_{-0.05} $ & $-12.23^{+0.05}_{-0.07} $ & $-11.96^{+0.05}_{-0.05} $ \\
		3C 207         & 0.684         & 4.27          & 2130          & $ 1.98E-01 $  & 0.62          & 0.86          & $1.59_{-0.23}^{+0.08} $ & $ 3.43_{-0.25}^{+14.98} $  & 0.96/150      & chi2          & $-12.02^{+0.13}_{-0.12} $ & $-11.79^{+0.19}_{-0.15} $ & $-11.58^{+0.15}_{-0.13} $ \\
		3C 208         & 1.109         & 3.12          & 9250          & $ 1.03E-01 $  &               &               & $1.63_{-0.14}^{+0.18} $ & $ 1.27_{-0.15}^{+0.22} $  & 0.70/14       & chi2          & $-12.44^{+0.04}_{-0.04} $ & $-12.24^{+0.10}_{-0.08} $ & $-12.03^{+0.07}_{-0.06} $ \\
		3C 212         & 1.049         & 3.79          & 434           & $ 4.17E-01 $  & 1.00          & 1.00          & $1.85_{-0.08}^{+0.03} $ & $ 3.65_{-0.20}^{+0.11} $  & 0.73/159      & chi2          & $-11.97^{+0.01}_{-0.01} $ & $-11.93^{+0.01}_{-0.01} $ & $-11.65^{+0.01}_{-0.01} $ \\
		3C 215         & 0.411         & 3.46          & 3054          & $ 4.58E-02 $  & 0.41          & 1.00          & $1.88_{-0.08}^{+0.09} $ & $ 6.86_{-0.98}^{+0.19} $  & 0.97/224      & chi2          & $-11.69^{+0.02}_{-0.02} $ & $-11.67^{+0.07}_{-0.05} $ & $-11.38^{+0.03}_{-0.03} $ \\
		3C 216         & 0.668         & 1.30          & 15002         & $ 2.17E-03 $  & 1.00          & 1.00          & $1.72_{-0.20}^{+0.26} $ & $ 2.16_{-0.25}^{+0.89} $  & 1.21/10       & chi2          & $-12.22^{+0.05}_{-0.05} $ & $-12.08^{+0.08}_{-0.09} $ & $-11.84^{+0.07}_{-0.05} $ \\
		3C 245         & 1.029         & 2.47          & 2136          & $ 0.00E+00 $  & 0.29          & 0.85          & $1.57_{-0.05}^{+0.10} $ & $ 2.28_{-0.06}^{+0.17} $  & 1.07/95       & chi2          & $-12.19^{+0.01}_{-0.01} $ & $-11.94^{+0.04}_{-0.03} $ & $-11.75^{+0.02}_{-0.02} $ \\
		3C 249.1       & 0.311         & 2.80          & 3986          & $ 0.00E+00 $  & 0.65          & 0.89          & $1.92_{-0.04}^{+0.13} $ & $ 6.61_{-1.09}^{+0.41} $  & 1.04/191      & chi2          & $-11.71^{+0.07}_{-0.07} $ & $-11.72^{+0.08}_{-0.08} $ & $-11.41^{+0.08}_{-0.07} $ \\
		3C 254         & 0.734         & 1.44          & 2209          & $ 3.66E-02 $  & 1.00          & 0.88          & $1.93_{-0.07}^{+0.10} $ & $ 2.83_{-0.10}^{+14.12} $  & 0.91/196      & chi2          & $-12.07^{+0.02}_{-0.02} $ & $-12.09^{+0.02}_{-0.02} $ & $-11.78^{+0.02}_{-0.02} $ \\
		3C 263         & 0.652         & 0.90          & 2126          & $ 0.00E+00 $  & 1.00          & 0.95          & $1.83_{-0.05}^{+0.09} $ & $ 2.97_{-0.05}^{+5.25} $  & 1.14/199      & chi2          & $-12.06^{+0.01}_{-0.01} $ & $-12.00^{+0.01}_{-0.01} $ & $-11.73^{+0.01}_{-0.01} $ \\
		3C 268.4       & 1.400         & 1.26          & 9325          & $ 3.88E-04 $  &               &               & $1.45_{-0.12}^{+0.17} $ & $ 1.17_{-0.10}^{+0.17} $  & 1.59/14       & chi2          & $-12.49^{+0.04}_{-0.04} $ & $-12.15^{+0.11}_{-0.09} $ & $-11.99^{+0.08}_{-0.07} $ \\
		3C 270.1       & 1.519         & 1.24          & 13906         & $ 1.69E-01 $  & 0.75          & 0.85          & $1.57_{-0.07}^{+0.07} $ & $ 1.43_{-0.05}^{+0.23} $  & 1.00/245      & chi2          & $-12.39^{+0.12}_{-0.11} $ & $-12.15^{+0.13}_{-0.13} $ & $-11.95^{+0.12}_{-0.13} $ \\
		3C 275.1       & 0.557         & 1.77          & 2096          & $ 8.56E-02 $  & 1.00          & 0.85          & $1.79_{-0.10}^{+0.11} $ & $ 2.41_{-0.07}^{+0.38} $  & 1.06/167      & chi2          & $-12.15^{+0.01}_{-0.01} $ & $-12.07^{+0.06}_{-0.06} $ & $-11.81^{+0.04}_{-0.03} $ \\
		3C 287         & 1.055         & 1.03          & 3103          & $ 1.02E-01 $  &               &               & $1.86_{-0.04}^{+0.04} $ & $ 1.25_{-0.04}^{+0.04} $  & 0.88/141      & chi2          & $-12.43^{+0.01}_{-0.01} $ & $-12.40^{+0.03}_{-0.03} $ & $-12.11^{+0.02}_{-0.02} $ \\
		3C 286         & 0.849         & 1.20          & 15006         & $ 6.36E-05 $  &               &               & $2.12_{-0.26}^{+0.51} $ & $ 1.10_{-0.14}^{+0.50} $  & 0.66/87       & cstat         & $-12.46^{+0.05}_{-0.04} $ & $-12.63^{+0.12}_{-0.10} $ & $-12.23^{+0.05}_{-0.05} $ \\
		3C 309.1       & 0.904         & 2.27          & 3105          & $ 0.00E+00 $  & 0.99          & 0.85          & $1.57_{-0.02}^{+0.07} $ & $ 3.95_{-0.05}^{+0.30} $  & 0.72/186      & chi2          & $-11.95^{+0.01}_{-0.01} $ & $-11.70^{+0.02}_{-0.01} $ & $-11.51^{+0.01}_{-0.01} $ \\
		3C 325         & 0.860         & 1.27          & 4818          & $ 3.01E+00 $  &               &               & $1.45_{-0.20}^{+0.21} $ & $ 0.29_{-0.06}^{+0.08} $  & 1.33/19       & chi2          & $-13.10^{+0.11}_{-0.11} $ & $-12.77^{+0.22}_{-0.15} $ & $-12.60^{+0.16}_{-0.13} $ \\
		3C 334         & 0.555         & 4.05          & 2097          & $ 8.38E-03 $  & 1.00          & 0.86          & $1.85_{-0.11}^{+0.18} $ & $ 3.09_{-0.16}^{+0.49} $  & 1.18/54       & chi2          & $-12.04^{+0.03}_{-0.03} $ & $-12.00^{+0.04}_{-0.04} $ & $-11.72^{+0.03}_{-0.03} $ \\
		3C 336         & 0.927         & 4.44          & 15008         & $ 2.66E-01 $  & 1.00          & 0.85          & $1.93_{-0.28}^{+0.24} $ & $ 2.16_{-0.34}^{+0.42} $  & 0.92/125      & cstat         & $-12.19^{+0.06}_{-0.06} $ & $-12.21^{+0.14}_{-0.11} $ & $-11.89^{+0.08}_{-0.06} $ \\
		3C 343         & 0.988         & 2.01          & 15011         & $ 5.04E-04 $  &               &               & $2.29_{-0.56}^{+0.82} $ & $ 0.04_{-0.01}^{+0.04} $  & 1.09/11       & cstat         & $-13.88^{+0.19}_{-0.15} $ & $-14.21^{+0.69}_{-0.26} $ & $-13.67^{+0.20}_{-0.15} $ \\
		3C 345         & 0.594         & 1.14          & 2143          & $ 0.00E+00 $  & 0.79          & 0.85          & $1.69_{-0.02}^{+0.09} $ & $ 8.79_{-0.10}^{+2.59} $  & 1.04/191      & chi2          & $-11.60^{+0.02}_{-0.02} $ & $-11.44^{+0.02}_{-0.02} $ & $-11.21^{+0.02}_{-0.02} $ \\
		3C 351         & 0.371         & 2.45          & 435           & $ 3.00E+00 $  & 0.23          & 0.85          & $1.90_{-0.13}^{+0.20} $ & $ 6.71_{-0.99}^{+2.07} $  & 1.05/112      & chi2          & $-11.70^{+0.09}_{-0.09} $ & $-11.71^{+0.17}_{-0.09} $ & $-11.38^{+0.10}_{-0.11} $ \\
		4C 16.49       & 1.296         & 6.63          & 9262          & $ 7.22E-04 $  &               &               & $1.83_{-0.12}^{+0.20} $ & $ 0.98_{-0.07}^{+0.16} $  & 0.87/115      & cstat         & $-12.54^{+0.04}_{-0.04} $ & $-12.49^{+0.10}_{-0.09} $ & $-12.21^{+0.06}_{-0.05} $ \\
		3C 380         & 0.691         & 5.78          & 3124          & $ 5.22E-02 $  & 0.95          & 0.85          & $1.76_{-0.10}^{+0.12} $ & $ 8.86_{-0.45}^{+4.01} $  & 1.04/130      & chi2          & $-11.59^{+0.01}_{-0.01} $ & $-11.48^{+0.02}_{-0.02} $ & $-11.23^{+0.01}_{-0.01} $ \\
		3C 432         & 1.805         & 5.65          & 5624          & $ 4.17E-01 $  &               &               & $1.76_{-0.10}^{+0.10} $ & $ 0.58_{-0.05}^{+0.05} $  & 1.14/40       & chi2          & $-12.77^{+0.04}_{-0.04} $ & $-12.66^{+0.08}_{-0.07} $ & $-12.41^{+0.05}_{-0.05} $ \\
		3C 454         & 1.757         & 5.26          & 21403         & $ 3.55E+00 $  &               &               & $1.75_{-0.19}^{+0.20} $ & $ 0.74_{-0.15}^{+0.20} $  & 0.80/23       & chi2          & $-12.67^{+0.10}_{-0.10} $ & $-12.56^{+0.19}_{-0.14} $ & $-12.31^{+0.13}_{-0.11} $ \\
		3C 454.3       & 0.859         & 6.63          & 4843          & $ 1.60E-01 $  & 0.37          & 0.99          & $1.67_{-0.05}^{+0.06} $ & $ 39.09_{-20.50}^{+29.26} $  & 0.94/387      & chi2          & $-10.93^{+0.27}_{-0.24} $ & $-10.74^{+0.27}_{-0.24} $ & $-10.53^{+0.28}_{-0.24} $ \\
		3C 455         & 0.543         & 4.32          & 15014         & $ 4.53E-03 $  &               &               & $1.65_{-0.21}^{+0.12} $ & $ 0.27_{-0.04}^{+0.04} $  & 0.78/110      & cstat         & $-13.11^{+0.06}_{-0.06} $ & $-12.92^{+0.19}_{-0.13} $ & $-12.70^{+0.12}_{-0.09} $ \\		
		\enddata
		\tablecomments{
			Column (1): 3CRR name; Column (2): Redshift; Column (3): Galactic neutral Hydrogen column density \citep{KalberlaEtal2005aap}, in units of $10^{20}~\rm cm^{-2}$; Column (4): Chandra observation ID; Column (5): Intrinsic Hydrogen column density, in units of $10^{22}~\rm cm^{-2}$; Columns (6-7): The alpha and f parameters in $ Sherpa $ for $jdpileup$ model \citep{DavisEtal2001apj}; Column (8): The power-law photon index and 1 $\sigma$ errors; Column (9): The normalization and 1 $\sigma$ errors of power-law component in $10^{-4}\rm~ photons\,keV^{-1}\,cm^{-2}\,s^{-1}$ at $ 1~\rm keV $; Column (10): Reduced $\chi^{2}$ and degree of freedom; Column (11): The statistical method, chi2 for $chi2xspecvar$, cstat for $cstat$; Columns ($ 12-14 $): The fluxes and 1 $\sigma$ errors after absorption corrections. 
		}		
	\end{deluxetable*}
\end{longrotatetable}

\subsection{Composite spectra}
\label{sec:3CRR_SED}

\begin{figure*}[htb!]
\includegraphics[width=0.5\textwidth]{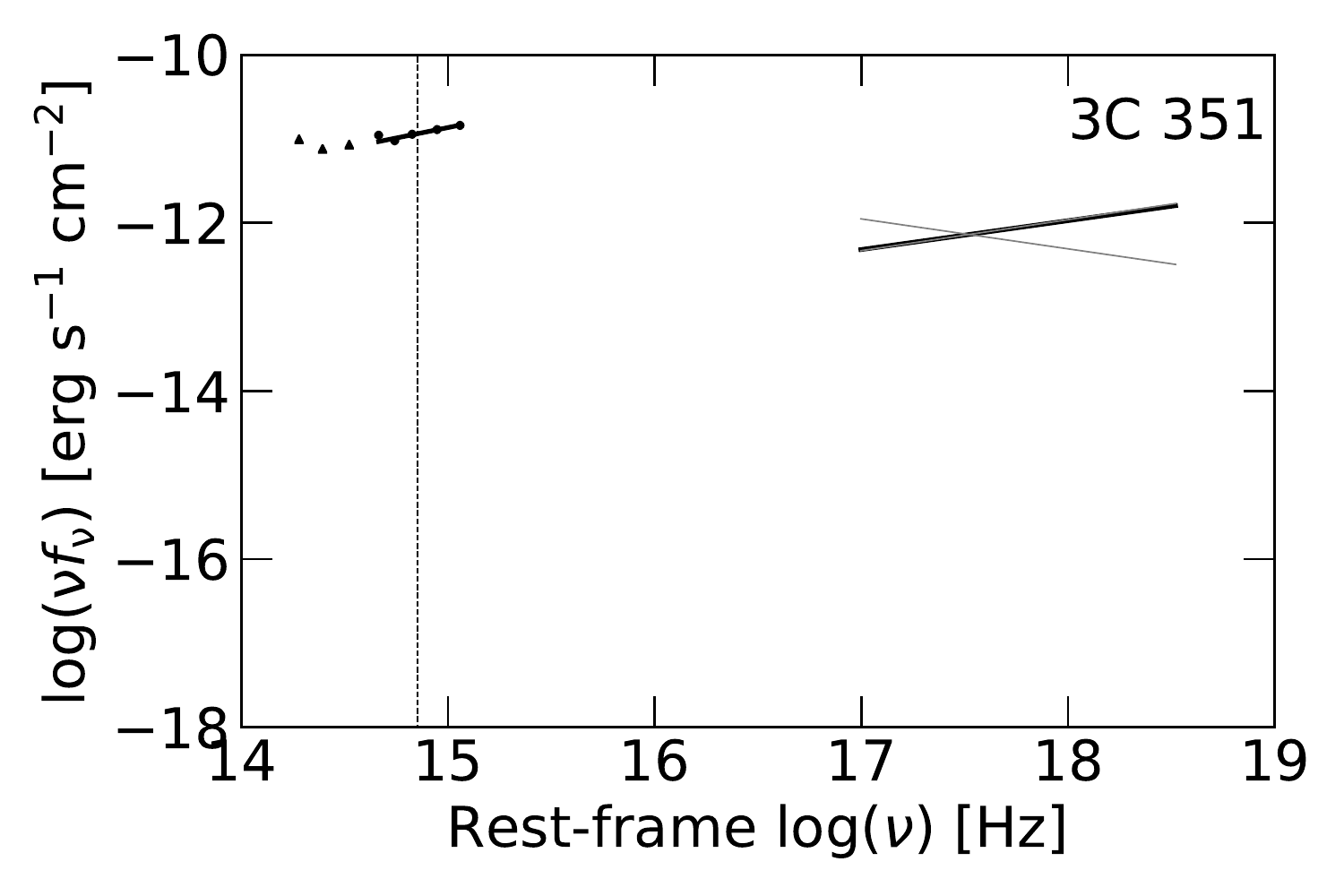}
\includegraphics[width=0.5\textwidth]{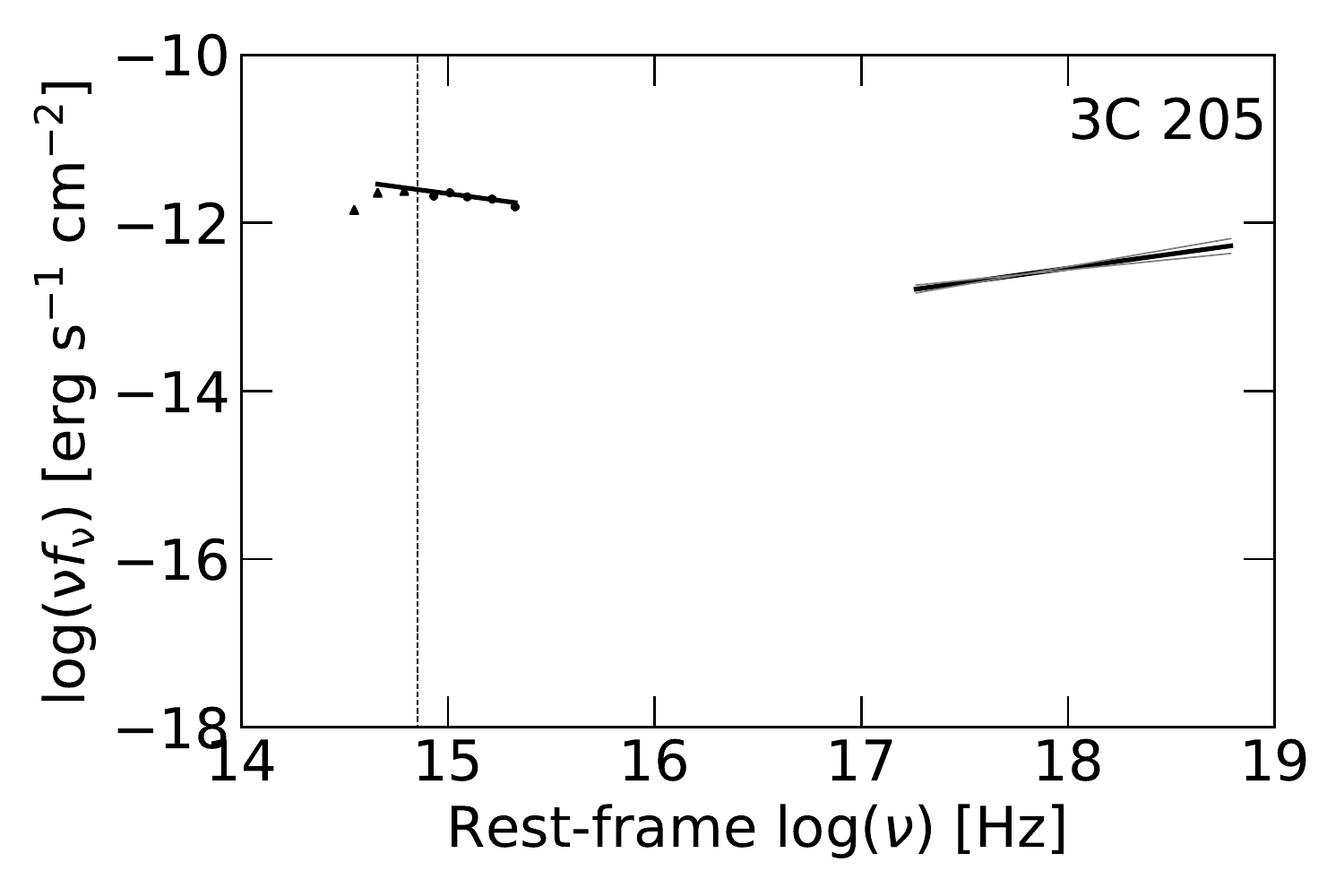}
\includegraphics[width=0.5\textwidth]{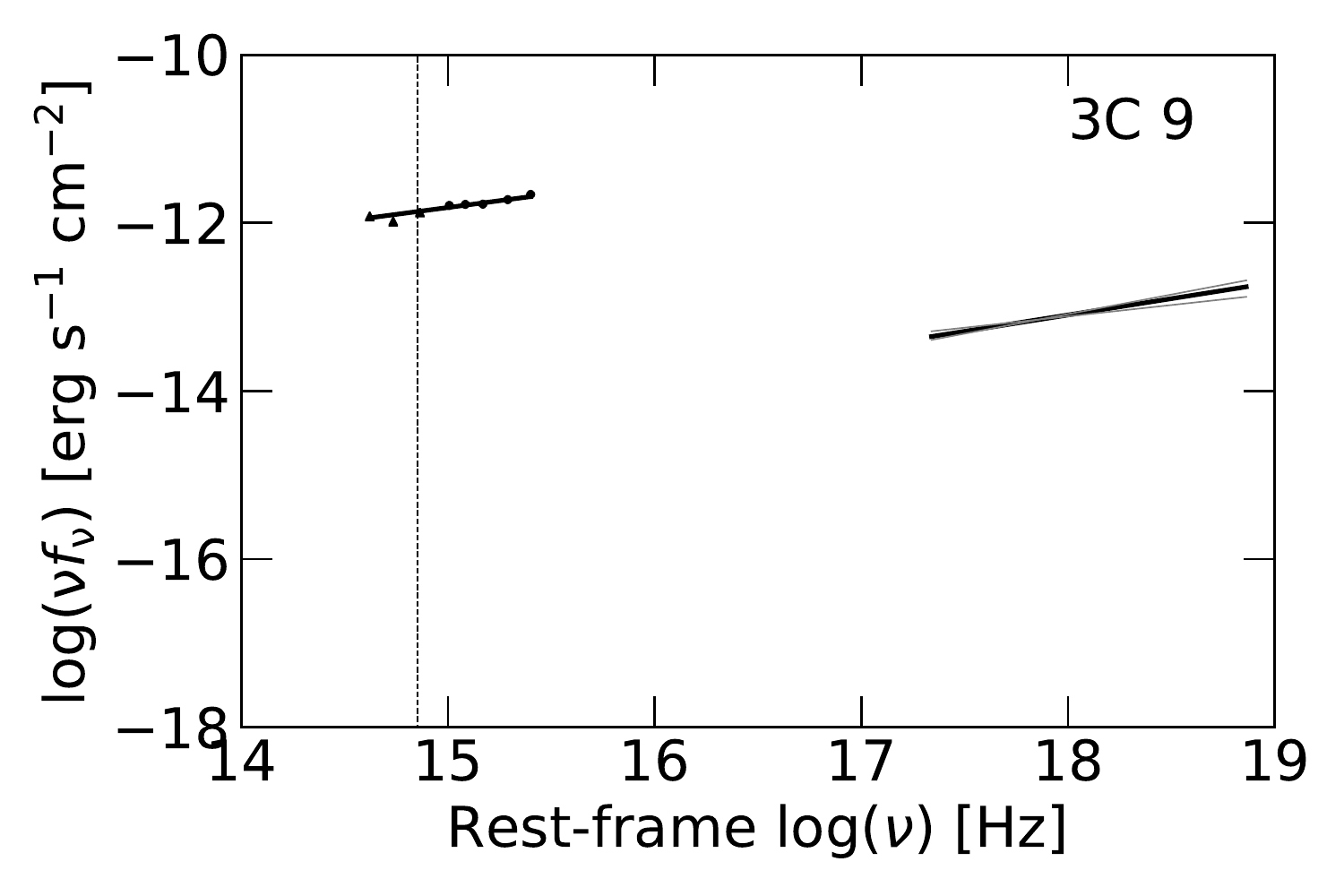}
\includegraphics[width=0.5\textwidth]{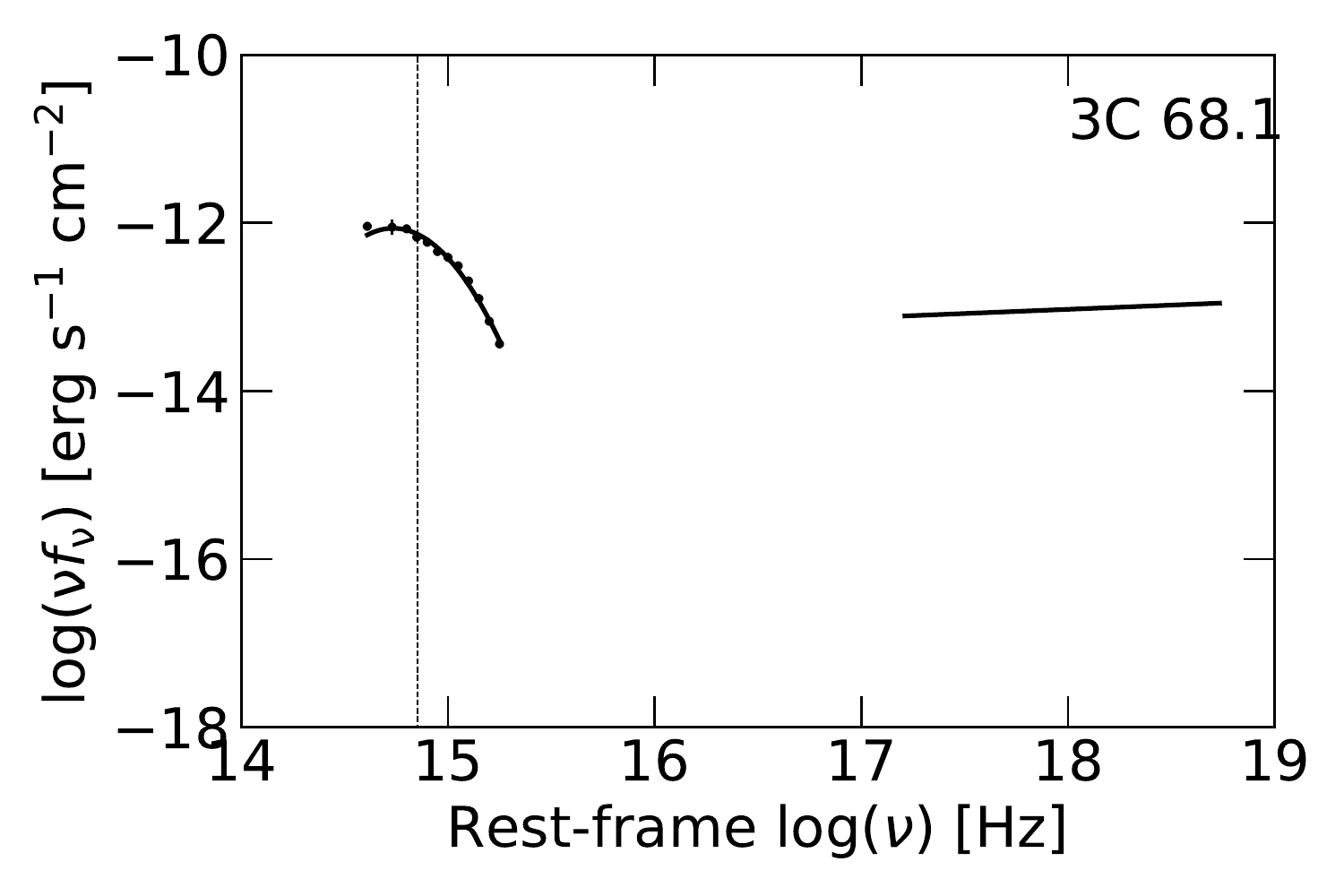}
\caption{The optical/IR and X-ray SED of four quasars, shown as examples for our sample. The power-law fits are shown in the plots, except for the log parabolic model for optical/IR data of 3C 68.1 (see text for details). The dashed line represents the position of normalization wavelength $4215~\rm \AA$.} 
\label{fig:3CRRSED}
\end{figure*}

As shown in S11, the composite spectrum constructed from the median values within frequency bins (e.g., S11) can be used as the representative of the overall SED for the studied sample. The median values at binned frequencies were obtained from the SED normalized to  a given optical frequency. For this reason, the optical/infrared SED will be needed in order to estimate the normalized optical flux. 
Therefore, to construct the composite X-ray spectrum for the sample, we firstly built the $\log(\nu{f}_\nu) - \log\nu$ SED at optical/infrared and X-ray bands of all quasars by directly calculating the rest-frame flux density $f_{\nu_{obs}(1+z)}$ from observational optical/infrared data $f_{\nu_{obs}}$ in combination with X-ray measurements. As examples, the SEDs of four quasars are shown in Figure \ref{fig:3CRRSED}. 

We followed the same method in S11 to construct the composite spectrum. The SED of individual objects was firstly normalized to rest-frame $4215~\rm \AA$. The flux density at rest-frame $4215~\rm \AA$ was directly estimated from the power-law fit on the optical/infrared data. 
When $4215~\rm \AA$ is covered by SDSS/NED/SSDC data, only these optical data were used in power-law fit. For those quasars at $0.6 <z\le 1.2$, the 2MASS J-band data was added in the fit. For the objects at $1.2 <z\le 2.0$, 2MASS J,H bands were added, and 2MASS J,H,K bands were used when $z>2.0$.
In all the cases except for 3C 68.1, the power-law model gave good fit to the continuum. The continuum of 3C 68.1 is convex, thus deviated from power-law (see Figure \ref{fig:3CRRSED}). This is likely caused by heavy extinction as shown in Appendix \ref{subsubsec:3c68}. Therefore, we fitted the continuum with the log-parabolic model \citep[$ f_\nu = k \nu^{(a+b\log\nu)} $,][]{2004A&A...413..489M}.

\begin{figure*}[tb]
	\centering
	\plotone{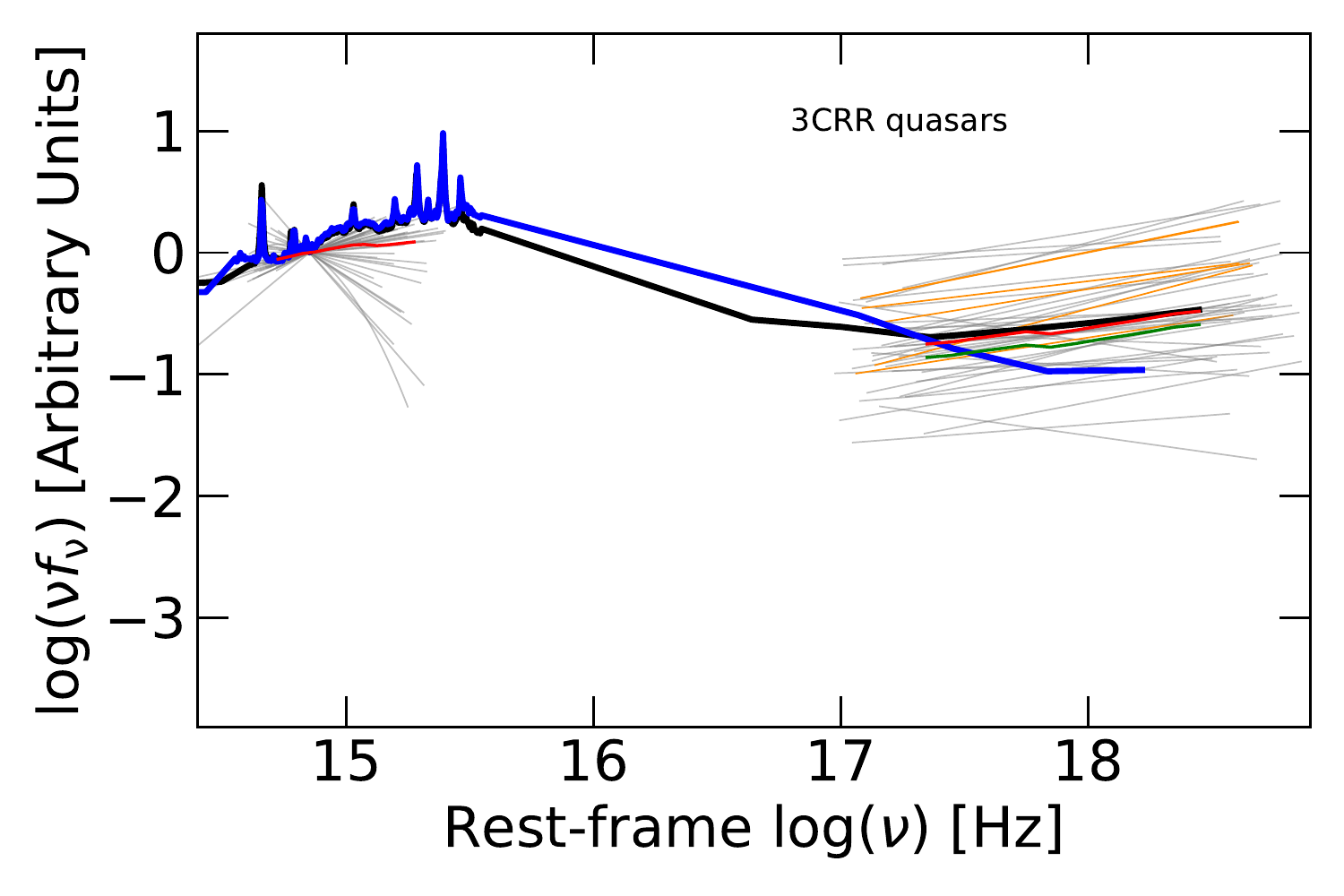}
	\plotone{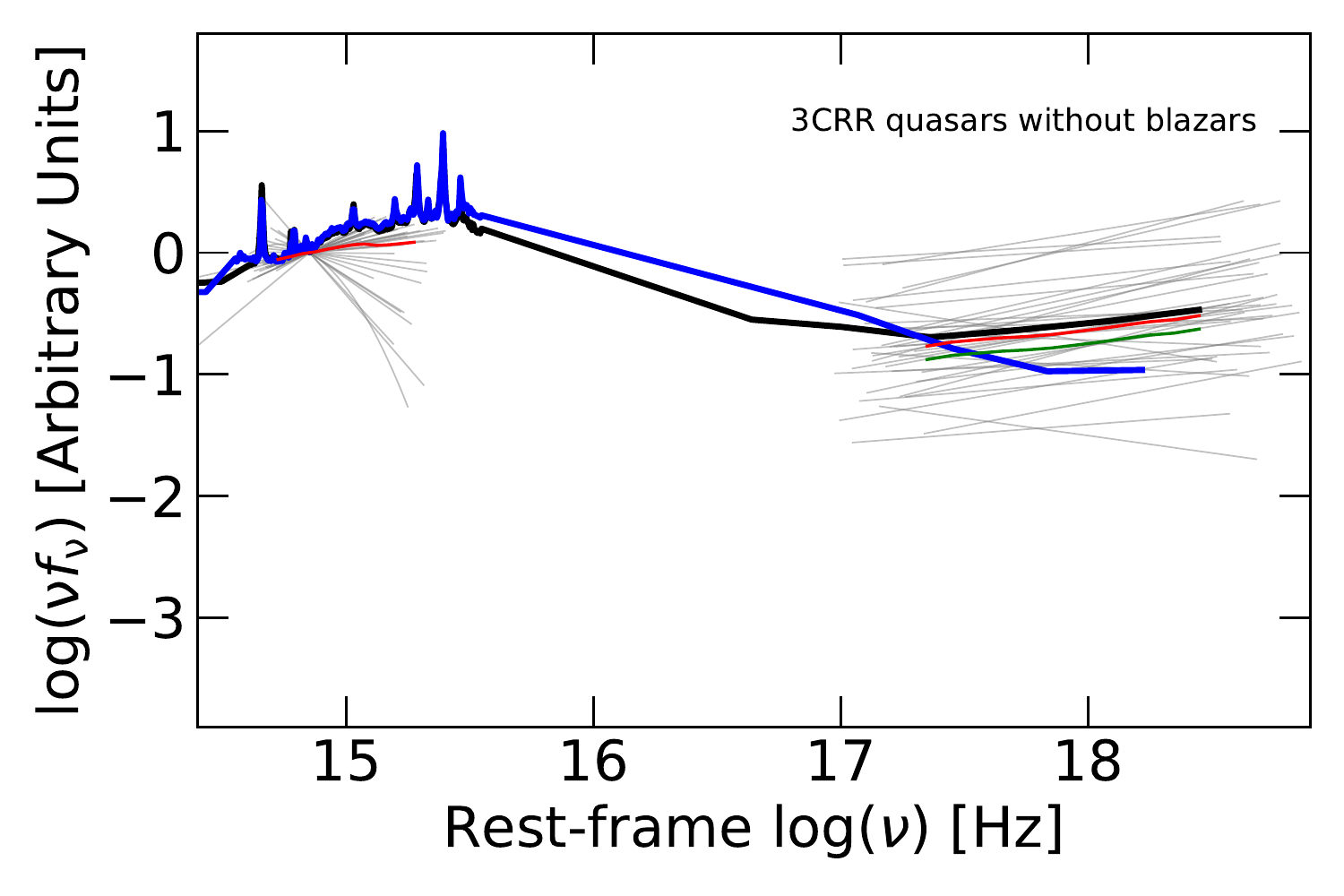}
	\caption{The median composite SED at optical and X-ray bands normalized at 4215$\rm \AA$ for our sample of 3CRR quasars, which are shown in red solid lines. The gray solid lines are normalized SED of individual objects. The thick black and blue solid lines are composite SEDs for RLQs and RQQs in \cite{ShangEtal2011apjs}, respectively. The green solid lines are composite X-ray spectra after extinction correction (see text for details). $Upper - $ for all 3CRR quasars, and the orange solid lines are X-ray spectra for six blazars; $Bottom - $ for 3CRR quasars excluding six blazars. \label{fig:SED}}
\end{figure*}

After normalization, the optical/IR continuum was rebinned with 12 bins in the $\log\nu$ range of 14.7 to 15.3 Hz. This frequency range was selected as it is covered by most quasars. At X-ray band, the spectra were resampled in the $\log\nu$ range of $17.45-18.45$ Hz, with a bin size of $\log\nu=0.1$. Following S11, the composite SED at optical/IR and X-ray band was constructed from the median values in each bin. 
Two composite median spectra were finally constructed, with one for all 43 3CRR quasars, and the other after excluding six blazars. The composite spectra are shown in Figure \ref{fig:SED} with red lines. 
it can be clearly seen that the composite X-ray spectrum of our 3CRR quasars is very close to the composite spectrum of RLQs in S11, however, it differs that of RQQs in S11 with flatter and stronger hard X-ray emission.

\section{Discussions}
\label{discussion}

\subsection{Selection effect}

There is a well-known relationship between $\alpha_{\rm ox}$ and $L_{2500\rm \AA}$ \cite[e.g.,][]{LussoEtal2010AandA}, in the way that more luminous sources have steeper optical to X-ray slope, which is defined as
\begin{equation}
\label{alphaox}
\alpha_{\rm ox} = -0.384\log(L_{2\rm keV}/L_{2500\rm \AA}).
\end{equation}
In this work, the composite X-ray spectrum is constructed using rest-frame $ 4215~\rm \AA$ as a reference, therefore, it will largely depend on the optical/UV luminosity. When comparing with other samples, like S11, this dependence needs to be investigated in order to avoid any selection effects. 

\begin{figure}[htb!]
	\includegraphics[width=1.0\columnwidth]{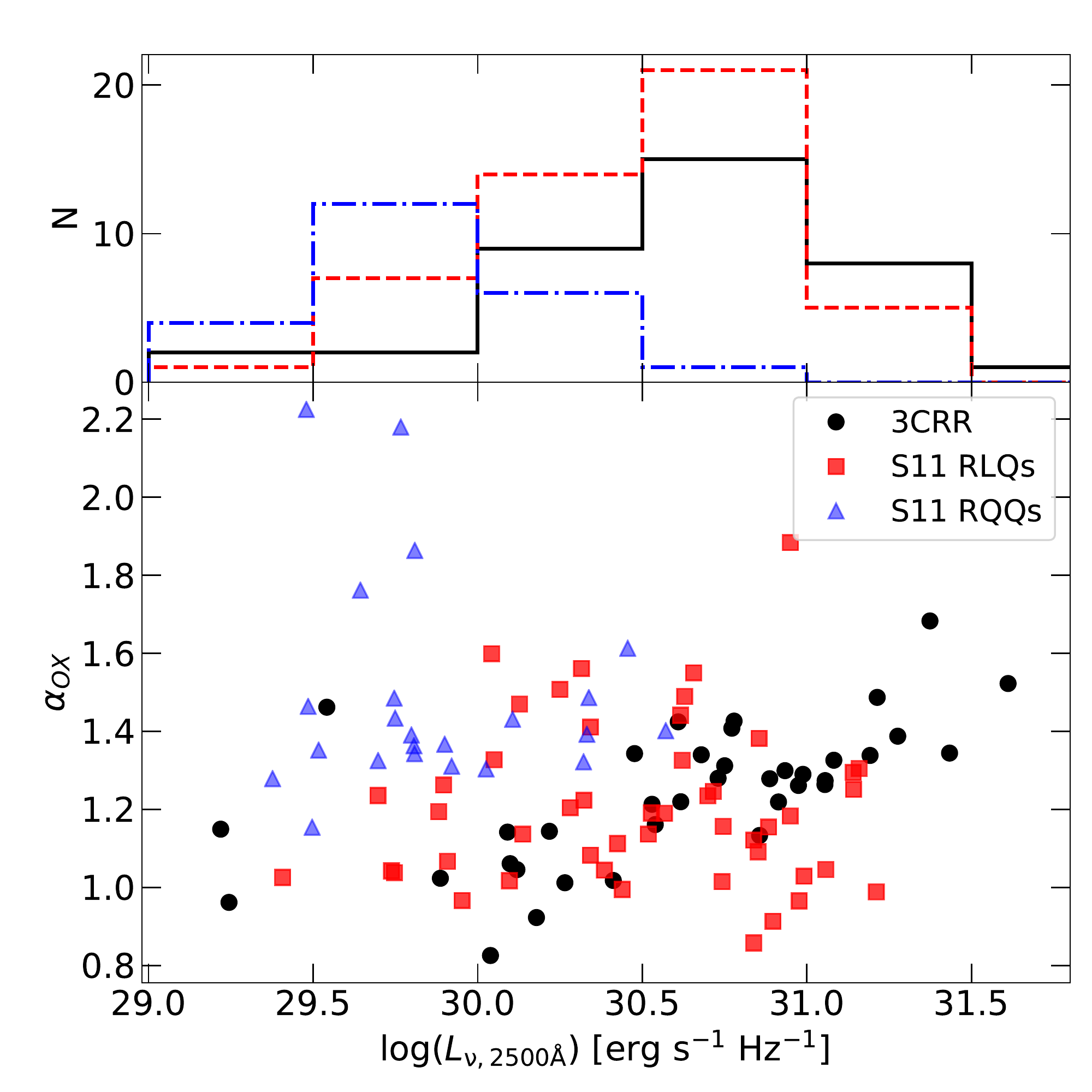}
	\caption{Relation between $\alpha_{\rm ox}$ and luminosity $L_{2500\rm \AA}$ for 3CRR and S11 quasars. The upper panel shows the histogram of luminosity $L_{2500\rm \AA}$. \label{fig:aox-lo} }
\end{figure}

In Figure \ref{fig:aox-lo}, we plotted the relationship between $\alpha_{\rm ox}$ and $L_{2500\rm \AA}$ for our sample and S11 sample. The luminosity at $2500~\rm \AA$ of S11 objects is taken from \cite{TangEtal2012ApJS}. 
As Figure \ref{fig:aox-lo} shows, 3CRR quasars and S11 RLQs have similar distribution on $\alpha_{\rm ox} - L_{2500\rm \AA}$ panel. However, S11 RQQs have larger $\alpha_{\rm ox}$ (mostly $>1.2$) than that of S11 RLQs and 3CRR quasars. While the $L_{2500\rm \AA}$ luminosity of 3CRR quasars are similar to those of S11 RLQs with Kolmogorov Smirnov (K$ - $S) test values $D=0.174$ and $P=0.511$, they are larger than those of S11 RQQs sample with $D=0.632$ and $P=9.536E-06$ from K$ - $S test. 

The similar composite X-ray spectrum of our sample with S11 RLQs seems to be reasonable considering their similar optical/UV luminosity. In contrast, the higher $L_{2500\rm \AA}$ for both samples than S11 RQQs, would imply lower X-ray emission if they follow the general $\alpha_{\rm ox} - L_{2500\rm \AA}$ relation \cite[e.g.,][]{Green1995ApJ,Anderson2003AJ,Vignali2003AJ}. However, this is opposite to our finding, i.e., higher X-ray emission in our sample compared to S11 RQQs. Therefore, the difference between RLQs and RQQs cannot be driven by selection effect related with optical/UV luminosity.

\begin{figure}[htb!]
	\includegraphics[width=1.0\columnwidth]{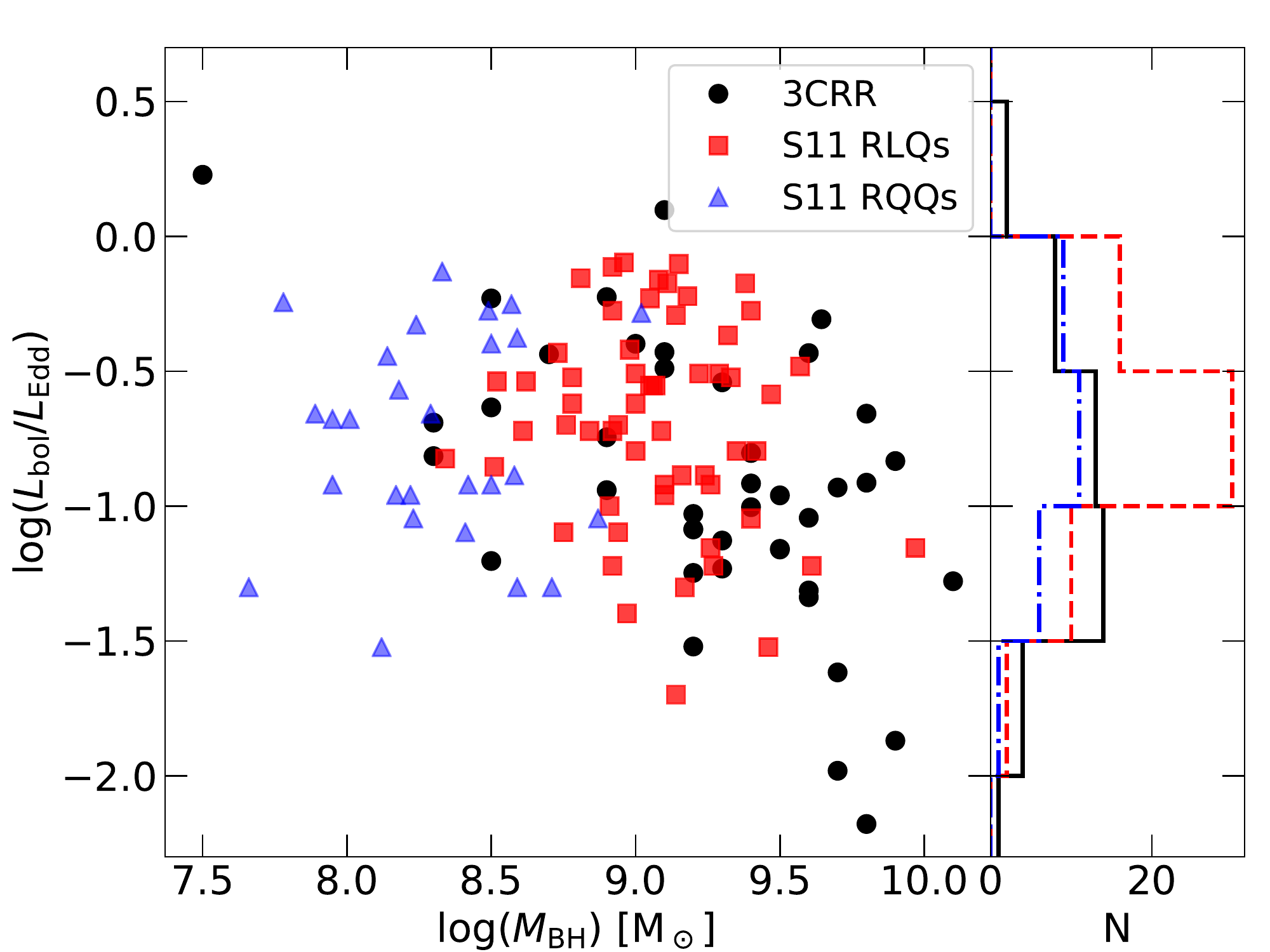}
	\caption{The distribution of black hole mass and Eddington ratio for 3CRR quasars and S11 samples. The right panel shows the histogram of Eddington ratio. \label{fig:bh-edd} }
\end{figure}

In addition to optical/UV luminosity, we compared black hole mass $M_{\rm BH}$ and Eddington ratio $L_{\rm bol}/L_{\rm Edd}$ of 3CRR quasars with S11 RLQs and RQQs, as shown in Figure \ref{fig:bh-edd}. The black hole mass and Eddington ratio of S11 quasars were directly taken from \cite{TangEtal2012ApJS}. 
The black hole masses of all 3CRR quasars were obtained from \cite{McLureEtal2006MNRAS}, except for 3C 216 and 3C 345 \citep{ShenEtal2011ApJS}, 3C 343 and 3C 455 \citep{Wu2009MNRAS}, and 3C 454.3 \citep{2001MNRAS.327.1111G}.
All these virial black hole masses were estimated with the empirical relationship between the broad line region radius and the optical/UV continuum luminosity, in combination with the line width of broad emission lines \citep[e.g.,][]{ShenEtal2011ApJS}. Depending on source redshift and availability of emission lines, various lines were used in the literature for our sample sources, with {H\,{\footnotesize $ \beta $}} usually at low-redshift sources, while {Mg\,{\footnotesize II}} or {C\,{\footnotesize IV}} at high redshift \citep[see e.g.,][]{ShenEtal2011ApJS}. 
We estimated the bolometric luminosity $L_{\rm bol}$ of 3CRR quasars with the relation established from S11 sample in \cite{RunnoeEtal2012MNRAS}, 
\begin{equation}
\label{bol_cor}
\log(L_{\rm iso}) = (4.89\pm1.66)+(0.91\pm0.04)\log(\lambda L_{\rm \lambda,5100\AA})
\end{equation}
The Eddington luminosity was calculated with black hole mass as $L_{\rm Edd}=1.25\times10^{38}({M_{\rm BH}}/{M_{\sun}})~\rm erg\ s^{-1}$ \citep{TangEtal2012ApJS}. 

We found from Figure \ref{fig:bh-edd} that the Eddington ratio of 3CRR quasars, S11 RLQs and RQQs are similar, although the black hole masses of 3CRR quasars and S11 RLQs are systematically larger than those of S11 RQQs. This implies a similar accretion mode in all three samples. Thus, it further indicates that the difference in composite X-ray spectrum between RLQs and RQQs is not caused by selecting different accretion systems, as manifested from the dependence of $\alpha_{\rm ox}$ on the Eddington ratio \cite[e.g.,][]{2019ApJ...883...76R}. 

\subsection{Extinction}
\label{sec:extinctedSED}

The advantage of selecting sources at radio band is that the radio emission is not subject to dust extinction, which is in contrast to optical selected sample. We found that the composite optical spectrum of 3CRR quasars is redder than S11 one (see Figure \ref{fig:SED}). Indeed, we found steep/red optical/UV spectra in many sources (e.g., 3C 14, 3C 68.1, 3C 190, 3C 205, 3C 212, 3C 216, 3C 268.4, 3C 270.1, 3C 325, 3C 343, 3C 345, 3C 454.3 and 3C 455), which can be seen in Figure \ref{fig:SED}. Prominent absorption lines were found in the optical spectra of many of these quasars, such as {Si\,{\footnotesize II}}, {Mg\,{\footnotesize II}}, {C\,{\footnotesize IV}}, {Fe\,{\footnotesize II}} etc. \cite[e.g.,][]{AldcroftEtal1994apjs}. It seems that there is some absorber locating at the line between source and observer. As the most extreme case, 3C 68.1 is likely a highly inclined and reddened quasar \citep{BoksenbergEtal1976apjl,BrothertonEtal1998apj} as mentioned earlier. 3C 325 is also a reddened quasar classified by \citet{GrimesEtal2005MNRAS}, and its spectrum shows lots of absorption lines in blue side. 3C 270.1 has a steep extreme ultraviolet (EUV) spectrum \citep{PunslyEtal2015MNRAS}. 

\begin{figure}[htb!]
	\includegraphics[width=1.0\columnwidth]{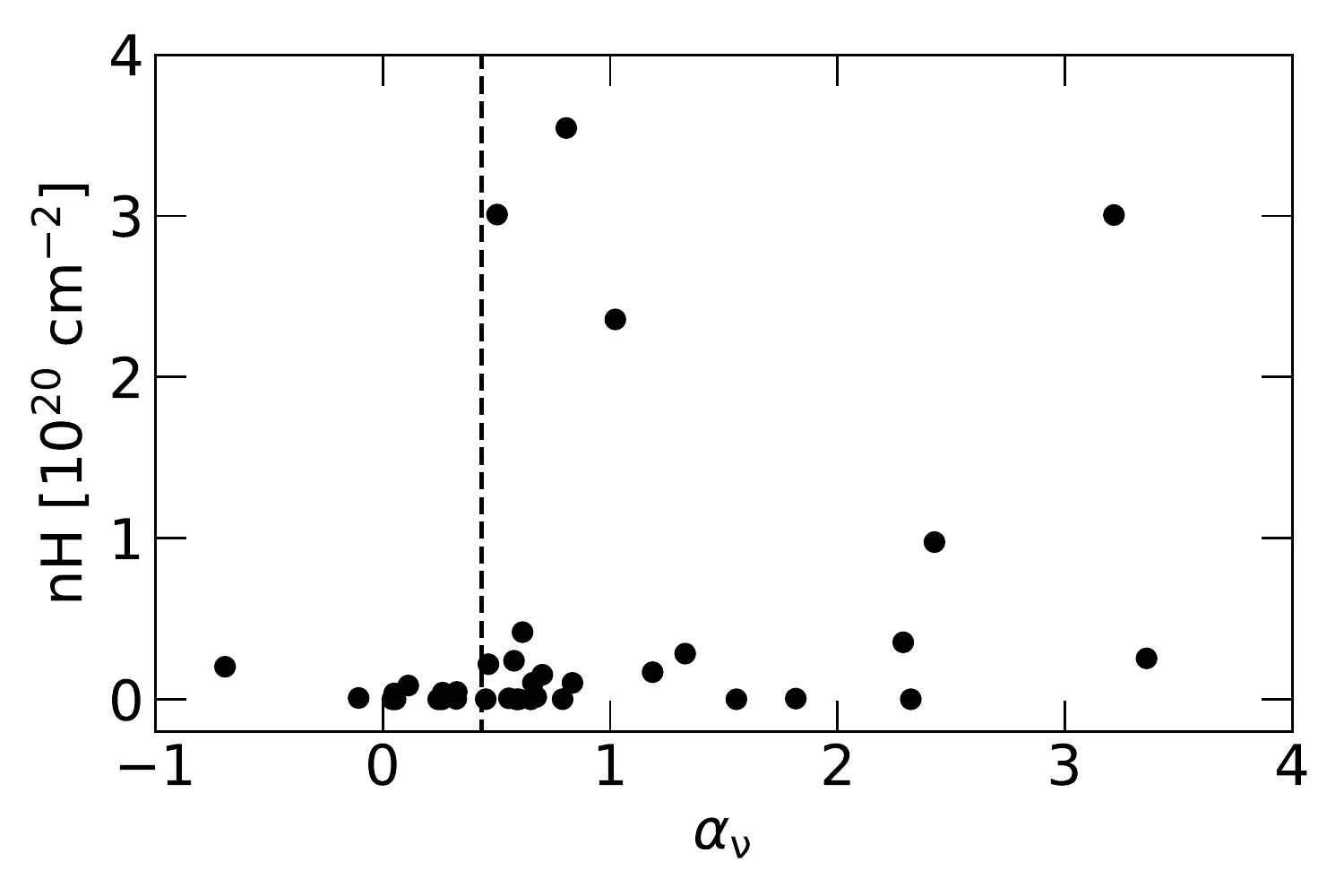}
	\caption{H\,{\footnotesize I} column density versus optical spectral index $\alpha_{\rm \nu}$. The dashed line indicates the spectral index of the composite spectrum of S11 RLQs ($\alpha_{\rm \nu} = 0.435$). \label{fig:alpha-nh} }
\end{figure}

In Figure \ref{fig:alpha-nh}, we plot the relation of intrinsic {H\,{\footnotesize I} column density ($\rm nH $, see Table \ref{table:3cfit1}) and optical spectral index ($ \alpha_{\rm \nu} $) of 3CRR quasars, where $\rm nH $ was calculated from X-ray spectral fitting. We found that the quasars with redder spectra tend to have relatively larger {H\,{\footnotesize I} column density, supporting dust extinction in some of these objects. If the emission at $4215~\rm \AA$ is indeed dust extincted, then the X-ray emission will be overestimated in the composite spectrum.

The systematic extinction of our 3CRR quasar sample can be evaluated by constructing the extinction curve by comparing our composite optical/UV spectrum with that of S11 RLQs sample. We normalized two composite spectra at $ \log \nu = 14.725 ~ \rm Hz $ (i.e., $\lambda = 5647.0~\rm \AA$, corresponding to the lowest frequency of our composite spectrum). We selected six line-free windows on the composite optical spectrum of S11 RLQs, then a power-law continuum is fitted from these windows (see Figure \ref{fig:ext_curve}). Using the continuum of S11 RLQs spectrum as reference, the extinction curve of our sample can be obtained, which is shown in Figure \ref{fig:ext_curve}. We found that the extinction curve is different from the Galactic reddening curve ($R_{\rm V}=3.1$) \citep{CardelliEtal1989IAUS}, SMC extinction curve, and the reddening curve of lobe-dominant radio AGNs in \cite{GaskellEtal2004ApJ}, with significantly higher value at short wavelength.
 		
The extinction at $4215~\rm \AA$, $A_{4215}$, is tentatively estimated by assuming a selective extinction $R_{\rm V}=3.1$. In this case, we get $A_{4215}=0.272$. Taking this extinction into account, the composite X-ray spectrum will shift down by about 0.11 dex. An adoption of $R_{\rm V}=5.3$ will result in a larger shift of about 0.18 dex. It can be seen from Figure \ref{fig:SED} that the X-ray composite spectra after extinction correction are lower than that of S11 RLQs, however the significant difference from S11 RQQs remains.  

\begin{figure}[tb]
	\includegraphics[width=1.0\columnwidth]{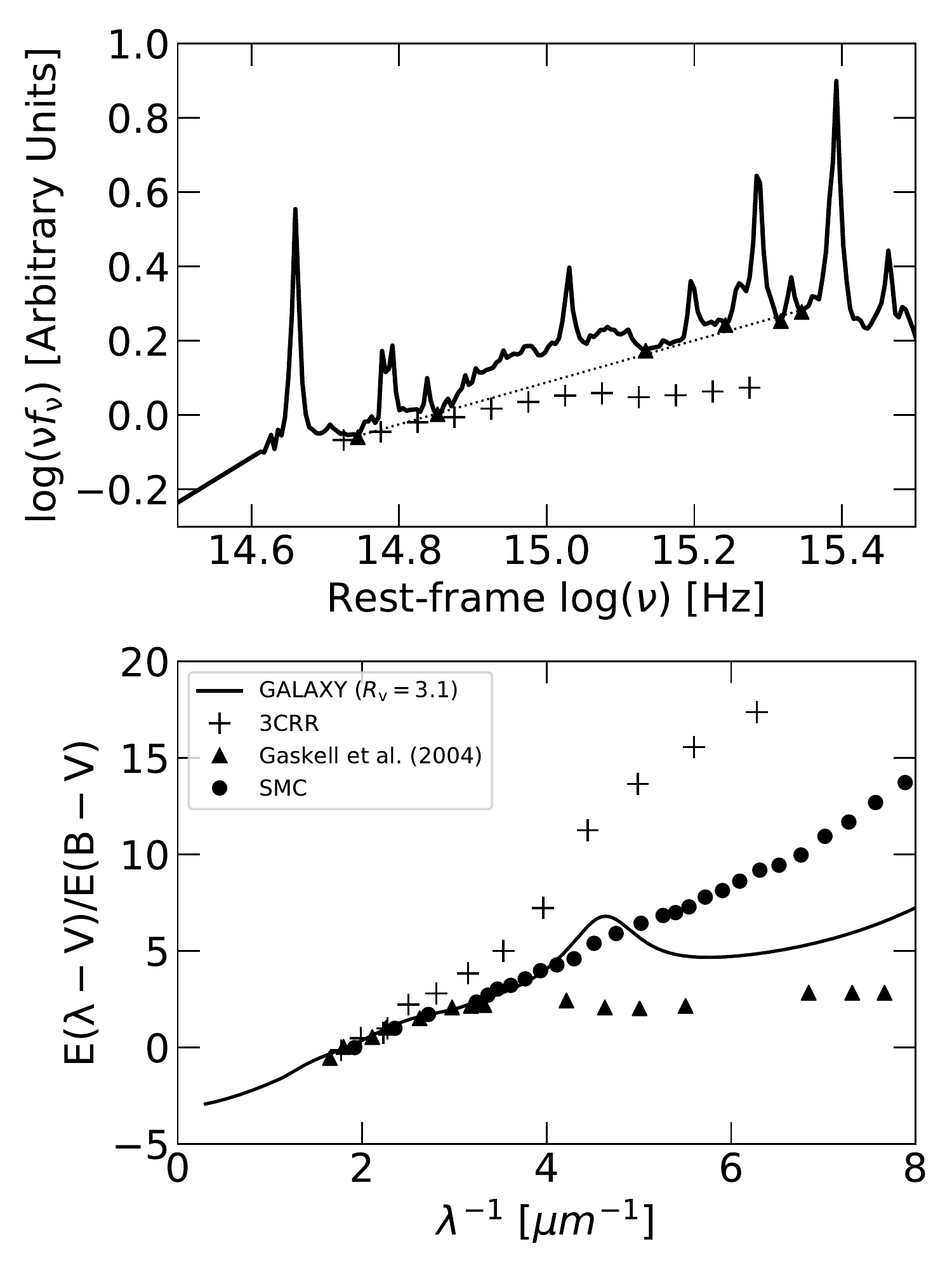}
	\caption{Extinction in 3CRR quasars: ($upper$) - The composite optical/UV spectrum of 3CRR quasars (black crosses) and S11 RLQs (black solid line). The dotted line is the power-law fit on the line-free windows (black triangles) on the S11 RLQs spectrum; ($bottom$) - Extinction curve of 3CRR quasars shown with black crosses. The filled triangles are from  the reddening curve of $ 0.1 \leq R_{c} < 1 $ quasars in \cite{GaskellEtal2004ApJ}, where $R_{c}$ is radio core-to-lobe flux ratio. The solid line is Galactic reddening curve of $R_V = 3.1$. The filled circles show the extinction curve of Small Magellanic Cloud \cite[SMC,][]{Prevot1984AandA}. \label{fig:ext_curve}}
\end{figure}

\subsection{Soft X-ray excess}
\label{sec:S11sed}

Compared to S11 RQQs, both composite X-ray spectra of our sample and S11 RLQs show lower soft X-ray emission. This difference may possibly be caused by the low fraction of soft X-ray excess detected in RLQs compared to RQQs \citep{Scott2011MNRAS,Scott2012MNRAS,Boissay2016AandA}. 

There are 27 RQQs in S11, of which 23 sources have available X-ray spectra, 7 from ROSAT and 16 from Chandra or XMM-Newton observations. As shown from the spectra fitting in S11, we found that all ROSAT data were fitted by a single power-law with a steep photon index ($\Gamma > 2.43$), and the rest quasars except for PG 0844+349 were fitted by a broken power-law model with a steep photon index at soft X-ray band as to ROSAT data and a flatter photon index at hard X-ray band. PG 0844+349 was fitted by a single power-law in S11, however by a double power-law model in \cite{Piconcelli2005AandA}. 
These results indicate that the soft X-ray excess is likely present in all S11 RQQs with a steeper photon index at soft X-ray band than hard X-ray band. 
Indeed, we found that RQQs commonly have soft X-ray excess, when we processed the updated Chandra and XMM-Newton data for S11 RQQs sample, which is now available for almost all sources. 

As mentioned in Section \ref{subsec:Xresults}, all X-ray spectra of 3CRR quasars except for 3C 351, can be well fitted by a single power-law. This seems to indicate that the soft X-ray excess is likely not evident for our present sample. However, it should be noticed that most of the Chandra observations used in this work arise from the 3CR snapshot survey \citep[][]{MassaroEtal2010ApJ, MassaroEtal2012ApJS, MassaroEtal2013apjs, 2018ApJS..234....7M, 2018ApJS..235...32S}. 
Therefore, the apparent non-detection of soft X-ray excess may be subject to the low photon counts, which precludes us to do spectral fit in great details.

\subsection{X-ray emission in RLQs}

It's well known that the difference of radio emission on the composite SED between RLQs and RQQs \cite[][S11]{ElvisEtal1994apjs} can be explained by the presence of powerful relativistic jets in RLQs, which is either weak or absent in RQQs \citep{2019NatAs...3..387P}. It is conceivable that the jet emission may contribute also at other bands than radio band \citep[e.g.,][]{2019ARA&A..57..467B}, especially in blazars when the relativistic jets is moving at small viewing angles \citep{Antonucci1993ARAA,UrryEtal1995PASP}. While the X-ray emission in RQQs may mainly be from the disk-corona system, the additional contribution from jets may present in RLQs \cite[e.g.,][]{WorrallEtal1987ApJ, MillerEtal2011ApJ, 2019MNRAS.490.3793L}. However, the fraction of X-ray emission in radio-loud AGNs that is from the jet, is a strongly debated issue \citep[e.g.,][]{GrandiEtal2004Sci, 2015ApJ...812...14M}. As illustrated in 3C 273, \cite{GrandiEtal2004Sci} found that the jet emission may contribute at both soft and hard X-ray bands based on the mixture model of Seyfert-like and jet-like emission on the BeppoSAX data, while the jet emission probably only presents at above several tens of keV based on broadband X-ray data, as found in \cite{2015ApJ...812...14M}.

\subsubsection{Comparison with previous works}

In Figure \ref{fig:SED}, while our composite X-ray spectrum is similar to that of S11 RLQs, we found that almost all six blazars (3C 207, 3C 216, 3C 309.1, 3C 380, 3C 345, and 3C 454.3) in our sample have higher X-ray flux than the composite spectrum, which may be caused by the significant jet contribution as the jet emission is usually thought to be highly boosted due to ``beaming effect''. However, although the X-ray composite spectrum of 3CRR quasars after excluding six blazars is lower than the original one, it is still close to that of S11 RLQs (see Figure \ref{fig:SED}). The student's $ t $-test shows a significant difference at $>99\%$ confidence level  between the distributions of normalized X-ray flux at $\log \nu = 18.22$ Hz of 3CRR quasars and S11 RQQs. As a result, the difference of composite X-ray spectrum between RLQs and RQQs is still significant. 

The more luminous X-ray in radio-loud AGNs compared to the radio-quiet population, has been also found in many works \cite[e.g.,][]{ZamoraniEtal1981ApJ, WorrallEtal1987ApJ, MillerEtal2011ApJ}. 
\cite{ZamoraniEtal1981ApJ} found that RLQs have more X-ray luminosity than RQQs at given optical luminosity. \cite{WorrallEtal1987ApJ} suggested that an ``extra" X-ray emission would dominate the observed X-ray flux in the majority of RLQs with flat radio spectra. 
\cite{MillerEtal2011ApJ} combined large, modern optical (e.g., SDSS) and radio (e.g., FIRST) surveys with archival X-ray data from Chandra, XMM-Newton, and ROSAT to generate an optically selected sample that includes 188 RIQs and 603 RLQs. The authors found that the excess X-ray luminosity compared to RQQs ranges from $\sim0.7$ to 2.8 for radio-intermediate quasars through the canonical $\sim3$ for RLQs to $>10$ for strongly radio-loud or luminous objects. Based on the results, they proposed a model in which the nuclear X-ray emission contains both disk/corona-linked and jet-linked components and demonstrated that the X-ray jet-linked emission is likely beamed but to a lesser degree than applies to the radio jet. 

On the other hand, \cite{1987ApJ...323..243W} found that the X-ray spectral slope of RLQs is flatter than that of RQQs, and argued that the ``two-components'' model (the flat and the steep components dominate the X-ray flux in RLQs and RQQs, respectively) can explain the result. On the contrary, \cite{1999ApJ...526...60S} provided evidence that radio-loud AGNs show comparable distribution of the X-ray continuum slope with radio-quiet AGNs.
Similarly, \cite{GuptaEtal2018MNRAS} found that radio galaxies are on average X-ray-louder than radio-quiet AGNs, however, the spectral slopes of two populations are very similar. The authors argued that in radio-loud and radio-quiet AGNs the hard X-rays are produced in the same region and by the same mechanism. The larger X-ray luminosities in radio-loud AGNs may result from larger radiative efficiencies of the innermost portions of the accretion flows around faster rotating black holes. More recently, \cite{2020MNRAS.492..315G} found that the average X-ray loudness of Type 1 and Type 2 radio-loud AGNs is very similar based on the sample selected from the $Swift$/BAT catalog. The X-ray loudness defined as the ratio of hard X-ray luminosity at 14-195 keV to MIR luminosity can be used to study the orientation-dependent X-ray emission since the MIR radiation is expected to be isotropic \cite[see e.g.,][]{2013ApJ...777...86L}. This similarity indicates negligible dependence of the observed X-ray luminosities on the inclination angle. Therefore, this seems to disfavor the significant jet emission at X-ray band as otherwise the jet emission will be expected to be more significant at smaller viewing angle, then higher X-ray loudness in Type 1 sources. 
As found in 3C 273, the well-known blazar, the model-fit on the X-ray spectrum from combined observations with Chandra, INTEGRAL, Suzaku, Swift, XMM-Newton, and NuSTAR observatories, gives that the coronal component is fit by $\Gamma_{\rm AGN} = 1.638\pm0.045$, cutoff energy $E_{\rm cutoff} = 47\pm15$ keV, and jet photon index by $\Gamma_{\rm jet} = 1.05\pm0.4$ \citep{2015ApJ...812...14M}. The beamed jet begins to dominate over emission from the inner accretion flow at $30-40$ keV.

The photon index at $2-10$ keV in our 3CRR quasars ranges from 1.45 to 2.32 with a median value of 1.69. This is in comparable with RQQs \cite[e.g.,][]{WangEtal2004ApJ,BrandtEtal2015A&ARv}, in good agreement with \cite{GuptaEtal2018MNRAS} that radio-loud and radio-quiet AGNs have similar distribution of hard X-ray photon index. 
A significant correlation between the X-ray photon index and the Eddington ratio has been found in radio-quiet AGNs in various works, in which the X-ray emission is thought to be from disk-corona system  \cite[e.g.,][]{WangEtal2004ApJ,BrandtEtal2015A&ARv}.  
Such correlation if found would be a clue on the emission mechanism of X-ray emission. 
We plot the relation of the X-ray photon index with the Eddington ratio for our 3CRR quasars sample in Figure \ref{fig:gamma-edd}. There is no significant correlation between the X-ray photon index and the Eddington ratio, in contrast to the finding for RQQs in \cite{WangEtal2004ApJ,BrandtEtal2015A&ARv}. A similar result has also been recently found in \cite{2019MNRAS.490.3793L} for a well-selected sample of radio-loud AGNs. 
This seems to imply that the X-ray emission may not be from, or at least not be dominated by the disk-corona system. In this case, it seems that the jet contribution can't be ignored. As a matter of fact, the photon index at $2-10$ keV of our sample is similar to that of FSRQs, which have an average value of $1.65\pm0.04$ \citep{DonatoEtal2001A&A}. However, the large uncertainties in the photon index, probably caused by low data quality, preclude us to draw a firm conclusion.

\begin{figure}[h]
	\includegraphics[width=1.0\columnwidth]{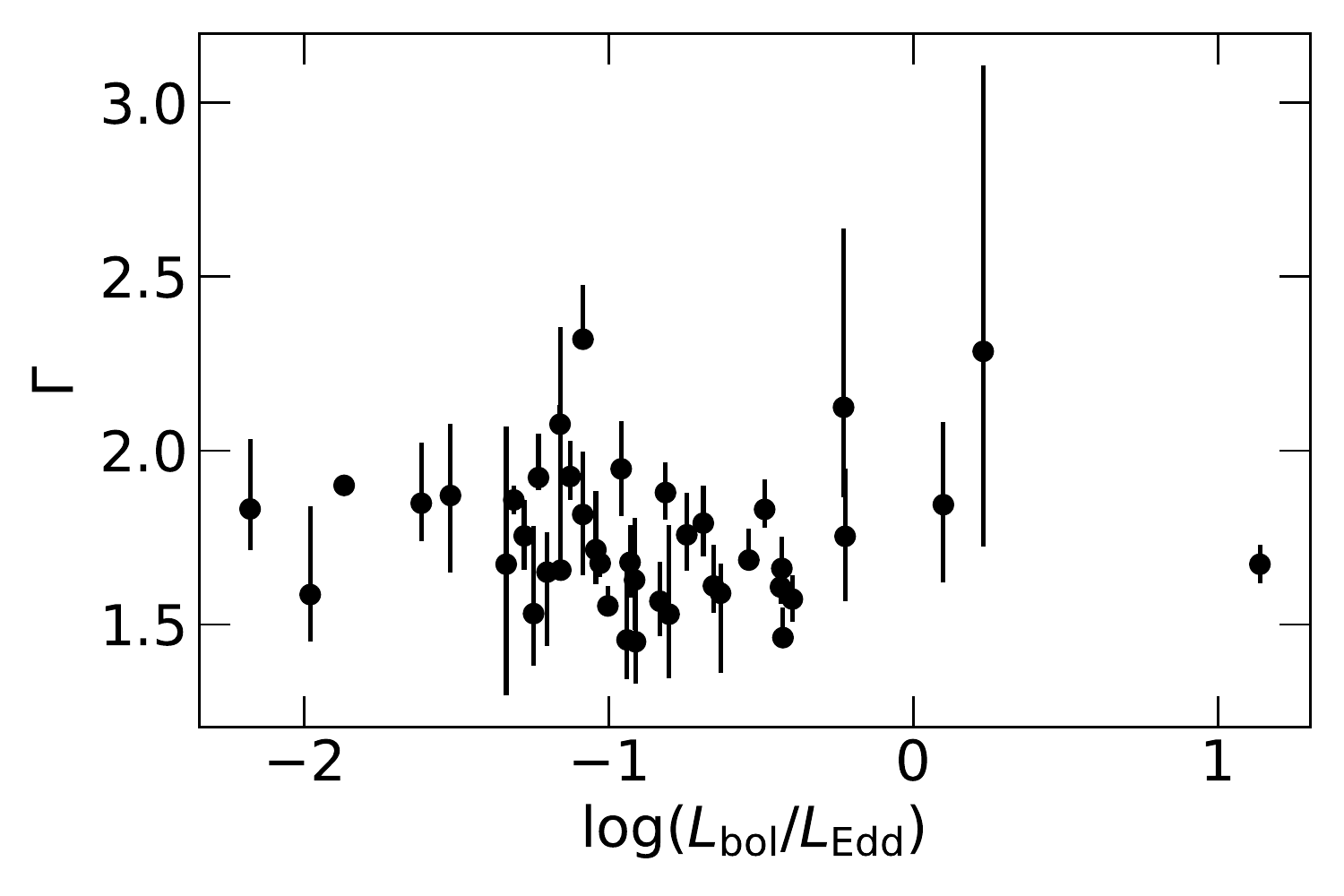}
	\caption{The relation between the X-ray photon index at $ 2-10~\rm keV $ and the Eddington ratio. \label{fig:gamma-edd} }
\end{figure}

\subsection{X-ray jets}
\label{sec:XJet}

In principle, the decomposition of X-ray emission into disk-corona and jet component can be performed when a high-quality X-ray spectrum is available at broad bands, as did in 3C 273 \citep{2015ApJ...812...14M}. However, this is usually hard for sample study due to the lack of broad band data. Although the fraction of X-ray emission in radio-loud AGNs that is from the jet is still unclear, it's contribution can be studied from the morphology of X-ray emission. Besides the unresolved central X-ray component, the X-ray emission can also be seen from the outer jet as shown in \cite{HarrisEtal2006ARAA}. The X-ray emission associated with radio jet knots has been studied in details for radio sources in 3CR catalog \citep{MassaroEtal2015ApJS}. 

XJET\footnote{\url{http://hea-www.harvard.edu/XJET/}} \citep{2010AIPC.1248..355H, 2010AIPC.1248..475M, 2011IAUS..275..160M, 2011ApJS..197...24M} collected 117 sources, which have extended X-ray emission associated with radio jets. After cross-matching 3CRR quasars with XJET catalog, we found 11 quasars have extended jet component in X-ray images, which has been studied in the literature, including 3C 9 \citep{FabianEtal2003MNRAS}, 3C47 \citep{HardcastleEtal2004ApJ}, 3C 207 \citep{BrunettiEtal2002aap}, 3C 212 \citep{AldcroftEtal2003ApJ}, 3C 254 \citep{DonahueEtal2003ApJ}, 3C 263 \citep{HardcastleEtal2002ApJ}, 3C 275.1 \citep{CrawfordEtal2003MNRAS}, 3C 345 \citep{SambrunaEtal2004ApJ}, 3C 351 \citep{BrunettiEtal2001ApJ}, 3C 380 \citep{MarshallEtal2005ApJS} and 3C 454.3 \citep{MarshallEtal2005ApJS}. Moreover, we checked the results of \cite{MassaroEtal2015ApJS}, and found that six more quasars (3C 181,  3C 191, 3C 215, 3C 245, 3C 325 and 3C 334) may have extended X-ray component related with jet knot, hotspot, or lobe. In addition, the extended X-ray emission from jet has also been found in 3C 270.1 \citep{WilkesEtal2012ApJ} and 3C 432 \citep{ErlundEtal2006MNRAS}. In total, there are 19 3CRR quasars having extended X-ray components associated with radio jets. 
 
\cite{2009ApJ...696..980M} reported the X-ray emission from the radio jet of 3C 17 with Chandra observations, and found that the high energy emission from the jet knots can be explained by both inverse Compton(IC)/cosmic microwave background (CMB) and Synchrotron process. Recently, \cite{2017MNRAS.470.2762M} found that the spectral fit of hotspots is consistent with the synchrotron emission, while the IC emission is found for lobes. In the case study of 3C 459, the extended X-ray emission can be well modelled by a plasma collisionally heated by jet-driven shocks \citep{2018A&A...619A..75M}. Based on the good correlation between the unabsorbed component of X-ray luminosity and
the 5-GHz core radio luminosity, \cite{HardcastleEtal2009MNRAS} argued that at least some, and in many cases all,
of the soft component of radio-source X-ray spectra originates in the jet is very hard to evade for a sample of 3CRR radio sources. As shown in \cite{MassaroEtal2015ApJS}, jet knot, hotspot, and lobe may all have X-ray emission, therefore, the emission from the jet components in the region of $2.5\arcsec$ radius will contribute in the composite spectrum constructed for our quasar sample. 
While the central $2.5\arcsec$ is unresolved in X-ray image, it can be usually resolved into core-jet structure with a central bright core and several jet components at radio band \cite[e.g., 3C 245, see Figure 1 in][]{MassaroEtal2015ApJS}. When the X-ray emission is extracted in central $2.5\arcsec$, the X-ray emission associated with jet components will naturally be included, although their flux fractions are largely unknown. 
As an additional check of jet emission at X-ray band, we studied the X-ay emission at different regions. However, no strong evidence of jet contribution was found due to the large uncertainties in the X-ray photon index (see Appendix \ref{sec:in-out}).

\section{Summary}
\label{summary}

In this work, we revisited the difference of composite X-ray spectrum between RLQs and RQQs, by using multi-band data for 3CRR quasars sample, which was selected at low radio frequency, thus less affected by the ``beaming effect''. We found that the composite X-ray spectrum of 3CRR quasars is similar to S11 RLQs after excluding blazars, still significantly different from RQQs. Although the photon index at $2-10$ keV of our sample is similar to RQQs, there is no strong correlation between the photon index and Eddington ratio, implying that other emission than the one from disk-corona system may also contribute at X-ray band, presumably from jet. 
The detection of X-ray emission from jet components has been reported in many sources from the literature, and the jet components might be included in extracting the X-ray spectrum from the central $2.5\arcsec$ region.
Our results suggest that the jet emission at X-ray band in RLQs could be related with the difference of composite X-ray
spectrum between RLQs and RQQs.

\acknowledgments

We thank the anonymous referee for valuable and insightful suggestions that improved the manuscript.
We thank Junxian Wang and Haritma Gaur for the help on X-ray data reduction, and Mai Liao, Muhammad Shahzad Anjum, Jiawen Li and Ye Yuan for useful discussions. This work is supported by the National Science Foundation of China (grants 11873073, U1531245, 11773056 and U1831138).

This research has made use of data obtained from the Chandra Data Archive and the Chandra Source Catalog, and software provided by the Chandra X-ray Center (CXC) in the application packages CIAO, ChIPS, and Sherpa.
This research has made use of the NASA/IPAC Extragalactic Database (NED) which is operated by the Jet Propulsion Laboratory, California Institute of Technology, under contract with the National Aeronautics and Space Administration.
Part of this work is based on archival data, software or online services provided by the Space Science Data Center - ASI.
This research has made use of the NASA/ IPAC Infrared Science Archive, 
which is operated by the Jet Propulsion Laboratory, California 
Institute of Technology, under contract with the National Aeronautics 
and Space Administration.
Funding for SDSS-III has been provided by the Alfred P. Sloan Foundation, the Participating Institutions, the National Science Foundation, and the U.S. Department of Energy Office of Science. The SDSS-III web site is http://www.sdss3.org/.
This publication makes use of data products from the Two Micron All Sky Survey, which is a joint project of the University of Massachusetts and the Infrared Processing and Analysis Center/California Institute of Technology, funded by the National Aeronautics and Space Administration and the National Science Foundation.

%

\vspace{5mm}
\facilities{SDSS, 2MASS, CXO}


\software{CIAO \citep{FruscioneEtal2006SPIE},
		  HEASoft ,
		  Python
          }



\appendix

\section{The X-ray spectral fit}
\subsection{3C 351}
\label{subsec:3c351fit}

3C 351 ($z=0.371$) has been observed by Chandra twice, a 9.15 ks observation with ID 435 on June 1, 2000, and a 50.92 ks observation with ID 2128 on August 24, 2001. These data were processed with CIAO as in Section \ref{subsec:xraydata}. The evt1 file counts per frame in $1.5\arcsec$ region are 0.142 and $0.327~\rm counts/frame$, for 435 and 2128 data, respectively. Therefore, the pile-up effect might be taken into account for the spectrum of 2128. We fitted these two spectra with absorbed power-law model and found that both spectra cannot be fitted well with significant excess at soft X-ray band (see the left panel in Figure \ref{fig:3c351}).

The soft X-ray spectrum of 3C 351 has been investigated by \cite{FioreEtal1993ApJ} using ROSAT data. The authors found that partial covering, soft excess, and warm absorber models can all fit the spectrum well, although only the warm absorber model (i.e., partially ionized absorbing material in the line of sight) gives a good $\chi^2$ for a typical value of high energy continuum slope. If the warm absorber model is correct, the strongest absorption edge lies in the range $0.58-0.76 ~\rm keV$, implying {O\,{\footnotesize IV}}$ - ${O\,{\footnotesize VII}} as the most likely absorbing ions. \cite{MathurEtal1994ApJ} provided strong evidence for the warm absorber model in the soft X-ray spectrum.

As did in \cite{FioreEtal1993ApJ}, 
we fitted the spectra of 3C 351 with the warm absorbed power-law model ($phabs*zxipcf*powerlaw$) using $chi2xspecvar$ statistical method. 
Here, we set the nH of photo-electric absorption ($ phabs $) to Galactic {H\,{\footnotesize I}} column density \citep{KalberlaEtal2005aap} and $zxipcf$ redshift to 0.371. We found two Chandra spectra can be well fitted by the model. As shown in the right panel of Figure \ref{fig:3c351}, the prominent absorption edge is visible below $1~\rm keV$. The results of spectral fitting are given in Table \ref{table:3c351fit}.

\begin{figure*}
	\includegraphics[width=0.5\textwidth]{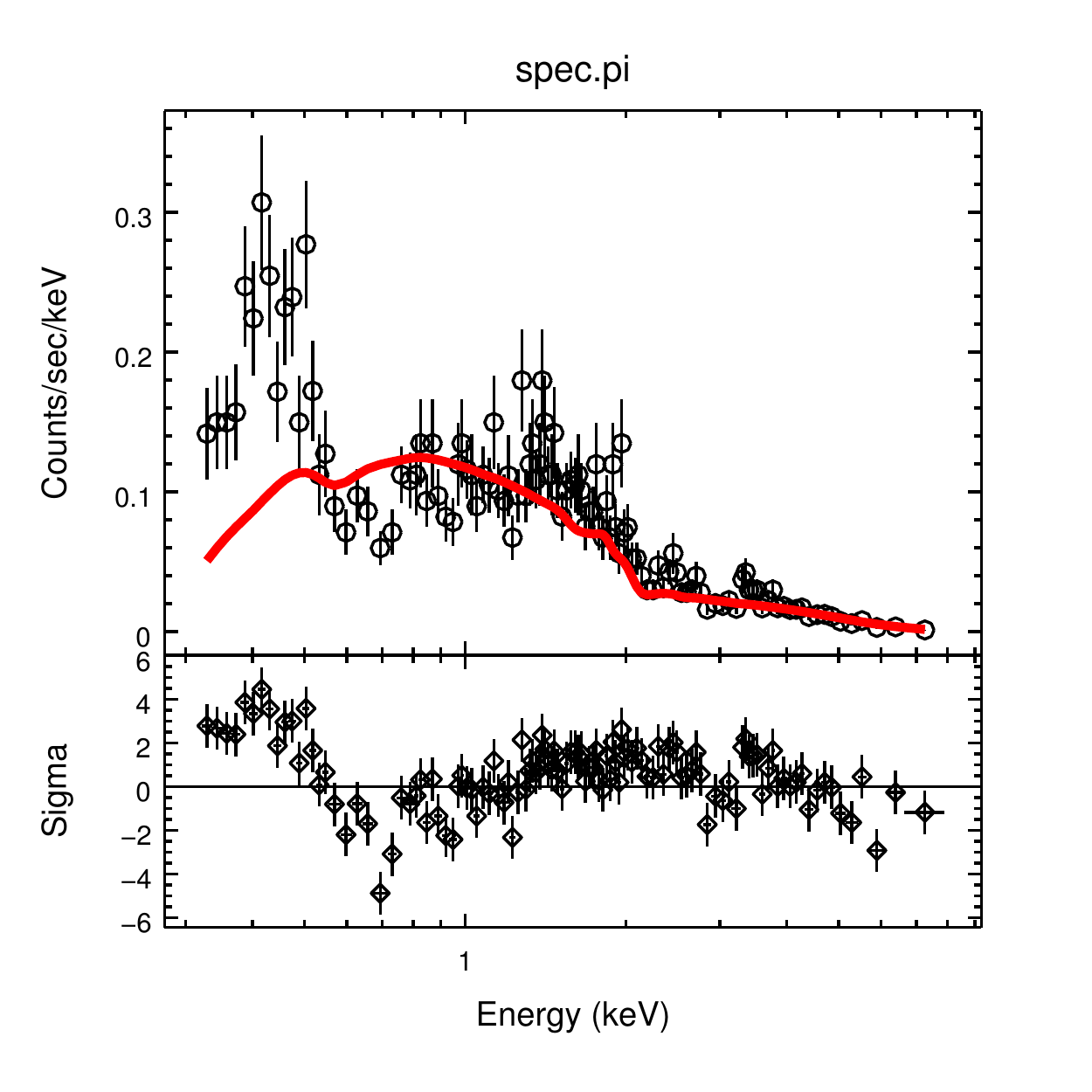}
	\includegraphics[width=0.5\textwidth]{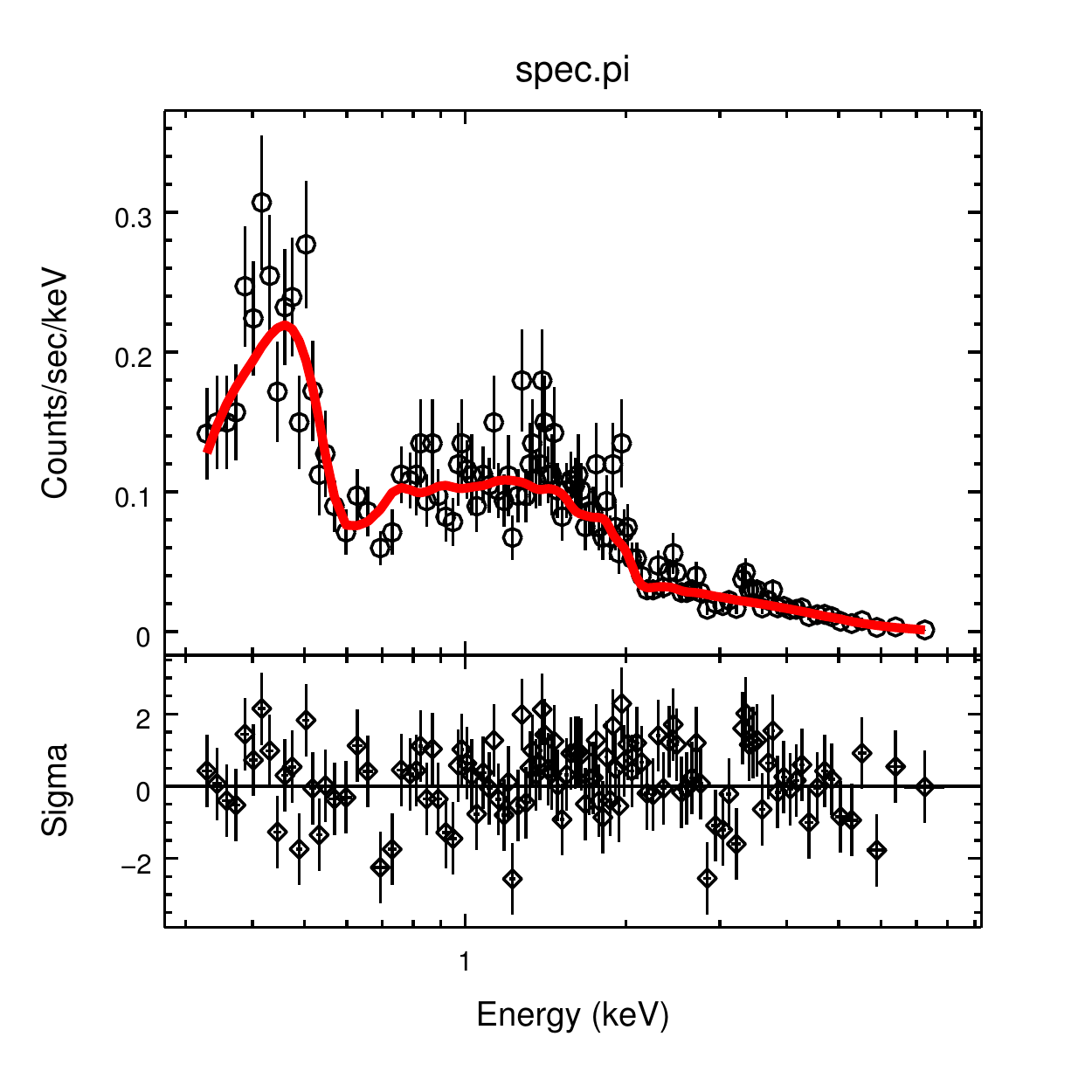}
	\caption{3C 351 spectrum of Chandra observation ID 435: ($Left$) - The spectrum is fitted by the absorbed power-law model ($phabs*zphabs*powerlaw$); ($Right$) - The spectrum is fitted with the warm absorbed power-law model ($phabs*zxipcf*powerlaw$). \label{fig:3c351}}
\end{figure*}

\begin{deluxetable}{rl|cc}
	\tablecaption{The spectral fitting for 3C 351\label{table:3c351fit}}
	\tablewidth{0pt}
	\tablehead{
		\colhead{ } & \colhead{ } & \colhead{$ 435 $} & \colhead{$ 2128 $} 
	}
	\decimalcolnumbers
	\startdata
	pile-up & alpha    &      &   0.42  \\
	& f      &     &   1.00 \\
	\hline
	zxipcf & nH($\rm 10^{22}~cm^{-2}$)  & 3.00  & 2.96  \\
	& $ \log(xi) $      & 1.90  & 1.20 \\
	& Covering Fraction   & 1.00  &  0.90 \\
	\hline
	Power-law & $ \Gamma $   & $ 1.66_{-0.03}^{+0.70} $  & $ 1.96_{-0.12}^{+0.13} $ \\
	& Norm. ($ 10^{-4} $)     & $ 4.54_{-0.12}^{+7.96} $ & $ 6.87_{-1.21}^{+0.56} $ \\
	& $\chi^2_{Red}/dof.$ & $ 1.10/114 $  &  $ 1.05/232 $  \\
	\hline
	Flux & $ 0.3-2.0~\rm keV $   & $ -11.89^{+0.01}_{-0.01} $ & $ -11.68^{+0.02}_{-0.02} $ \\
	& $ 2.0-10.0~\rm keV $   & $ -11.70^{+0.02}_{-0.02} $ & $ -11.72^{+0.04}_{-0.04} $ \\
	& $ 0.3-10.0~\rm keV $   & $ -11.49^{+0.01}_{-0.01} $ & $ -11.40^{+0.02}_{-0.02} $ \\
	\enddata
	\tablecomments{Rows 1 and 2: pile-up parameters; Rows $ 3-5 $: $zxipcf$ parameters; Rows $ 6-8 $: power-law parameters, where $\Gamma$ is photon index, Norm. is $\rm photons\,keV^{-1}\,cm^{-2}\,s^{-1}$ at $1~\rm keV$, and $\chi^2_{Red}/dof.$ is reduced $\chi^{2}$ and degree of freedom; Rows $ 9-11 $: the absorption-corrected flux ($\log f$ in units of $ \rm erg~cm^{-2}~s^{-1} $).}
\end{deluxetable}%

\subsection{3C 68.1}
\label{subsubsec:3c68}

3C 68.1 is an optically red quasar with very steep optical spectrum ($\alpha\sim6.1$) \citep{BoksenbergEtal1976apjl}, and is also highly polarized at $3000-8000~\rm \AA$ \citep{MooreEtal1981ApJ,MooreEtal1984ApJ}. \cite{BrothertonEtal1998apj} argued that 3C 68.1 is a highly inclined quasar and shows intrinsic ultraviolet absorption lines, like, {Mg\,{\footnotesize II}}, and {Ca\,{\footnotesize II}\,K}. 3C 68.1 probably has absorbed X-ray emission \citep{BregmanEtal1985ApJ}. All these observational features indicate that 3C 68.1 is likely an edge-on system along a line of sight through dusty, and ionized gas, perhaps part of torus \citep{BrothertonEtal2002AJ}.

3C 68.1 has been observed with an exposure time of $3.05~\rm ks$ by Chandra on February 10, 2008 (ID: 9244). 
The X-ray spectrum extracted from $2.5\arcsec$ circle was fitted with an absorbed power-law model with fixed Galactic absorption ($phabs*zphabs*powerlaw$). An extremely flat spectrum was found with $\Gamma = 0.46_{-0.33}^{+0.41}$. The flat X-ray spectrum is consistent with the finding in \cite{GouldingEtal2018ApJ} for extremely red quasars, and it may imply severe absorption in the source as shown from the extremely red optical spectrum and intrinsic ultraviolet absorption. Therefore, we fitted the X-ray spectrum again by fixing $\Gamma = 1.9$, and found the intrinsic X-ray absorption of $N_H \sim 5.31 \times 10^{22} ~\rm cm^{-2}$.

\section{X-ray emission at different regions}
\label{sec:in-out}

To further get the clues of jet contribution at X-ray band, we tried to compare the X-ray spectra extracted from different regions, $0.5-2.0\arcsec$ annulus and central $1.5\arcsec$, in which the jet contribution is expected to be different.  
3C 68.1 and 3C 343 are excluded due to low photon counts in $0.5-2.0\arcsec$ annulus. We found that all Chandra X-ray spectra can be well fitted by absorbed power-law model except for 3C 351 (see Table \ref{table:3cfit2}). 

The distribution of X-ray photon index at $2-10$ keV from $0.5-2.0\arcsec$, $1.5\arcsec$ and $2.5\arcsec$ are compared in Figure \ref{fig:gammahist}. While the photon index of $1.5\arcsec$ is similar to that of $2.5\arcsec$, we found that most sources (34/41) have flatter spectra from outer region ($0.5-2.0\arcsec$ annulus) than inner region ($1.5\arcsec$ circle). While this may be caused by the intrinsic difference in the emission at different regions, the effect that Chandra/ACIS detector has broader PSF at hard X-ray can't be ignored. To examine the PSF effect, we used MARX simulate events on the X-ray data of our sample. After excluding the sources with pile-up effect, and low photon counts data, we extracted the X-ray spectra of 26 quasars from the source-centered region with $5\arcsec$ radius out of background region ($5-20\arcsec$ annulus) and fitted them with 1-D polynomial function ($polynom1d$), which were used as input spectra to MARX. The X-ray spectra of source-centered $1.5\arcsec$ circle and $0.5-2.0\arcsec$ annulus extracted from simulated data were fitted with an absorbed power-law model ($phabs*powerlaw$). 
We found the X-ray spectra from outer region are flatter than those from inner region, with median photon index of 1.49 and 1.68 for the former and the latter, respectively. The K$ - $S test shows significantly different distributions of X-ray photon index in two regions with $ D=0.440 $ and $ P=0.010 $.
The resulting photon index at $2-10~\rm keV$ is shown in Figure \ref{fig:spec-gamma}. Indeed, the PSF effect results in flatter spectra in outer region than inner region, which even is more significant than the observational result. Therefore, by taking this PSF effect into account, the X-ray spectra at outer region is slightly steeper than inner region. However, the firm conclusion can't be drawn due to the large uncertainties in the photon index. 

\begin{deluxetable*}{lrr|rrr|rll}
	\tablecaption{The X-ray spectral fitting for different regions \label{table:3cfit2}}
	\tablewidth{0pt}
	\tablehead{
		\multicolumn{3}{c}{ } & \multicolumn{3}{c}{$ 1.5\arcsec $ circle} & \multicolumn{3}{c}{$ 0.5-2.0\arcsec $ annulus} \\
		\colhead{Name} & \colhead{CXO ID} & \colhead{$\Gamma_0$} & \colhead{$\Gamma_1$} &
		\colhead{$\chi^2_{Red}/dof.$} & \colhead{Stat.} & \colhead{$\Gamma_2$} & \colhead{$\chi^2_{Red}/dof.$} &
		\colhead{Stat.} 
	}
	\decimalcolnumbers
	\startdata
    3C 9        & 1595          & $1.61_{-0.08}^{+0.12}$ & $1.62_{-0.10}^{+0.12}$ &  1.29/27         &  chi2            & $1.55_{-0.16}^{+0.17}$ &  0.90/111        &  cstat \\
	3C 14       & 9242          & $1.53_{-0.18}^{+0.26}$ & $1.43_{-0.11}^{+0.22}$ &  0.90/11         &  chi2            & $1.31_{-0.29}^{+0.31}$ &  0.95/52         &  cstat \\
	3C 43       & 9324          & $1.53_{-0.15}^{+0.25}$ & $1.48_{-0.14}^{+0.27}$ &  0.95/110        &  cstat           & $1.65_{-0.31}^{+0.27}$ &  1.07/34         &  cstat \\
	3C 47       & 2129          & $1.87_{-0.22}^{+0.21}$ & $1.76_{-0.20}^{+0.22}$ &  1.08/282        &  chi2            & $1.32_{-0.04}^{+0.06}$ &  1.13/128        &  chi2 \\
	3C 48       & 3097          & $2.32_{-0.01}^{+0.16}$ & $2.31_{-0.01}^{+0.17}$ &  1.24/170        &  chi2            & $2.21_{-0.03}^{+0.07}$ &  0.84/94         &  chi2 \\
	3C 138      & 14996         & $1.46_{-0.11}^{+0.24}$ & $1.59_{-0.10}^{+0.12}$ &  1.05/205        &  cstat           & $1.46_{-0.11}^{+0.33}$ &  0.89/97         &  cstat \\
	3C 147      & 14997         & $1.85_{-0.22}^{+0.24}$ & $1.83_{-0.22}^{+0.24}$ &  0.75/106        &  cstat           & $1.40_{-0.22}^{+0.40}$ &  1.00/38         &  cstat \\
	3C 175      & 14999         & $1.57_{-0.10}^{+0.11}$ & $1.51_{-0.11}^{+0.13}$ &  0.57/18        &  chi2           & $1.55_{-0.14}^{+0.37}$ &  0.86/85         &  cstat \\
	3C 181      & 9246          & $1.72_{-0.10}^{+0.17}$ & $1.70_{-0.10}^{+0.15}$ &  1.14/116        &  cstat           & $1.66_{-0.20}^{+0.26}$ &  1.13/49         &  cstat \\
	3C 186      & 9774          & $2.08_{-0.03}^{+0.06}$ & $2.07_{-0.04}^{+0.02}$ &  0.99/165        &  chi2            & $1.94_{-0.07}^{+0.04}$ &  1.04/88         &  chi2 \\
	3C 190      & 17107         & $1.61_{-0.05}^{+0.06}$ & $1.60_{-0.06}^{+0.07}$ &  1.17/79         &  chi2           & $1.62_{-0.06}^{+0.15}$ &  0.81/33         &  chi2 \\
	3C 191      & 5626          & $1.68_{-0.10}^{+0.11}$ & $1.68_{-0.10}^{+0.11}$ &  0.98/36         &  chi2            & $1.57_{-0.11}^{+0.14}$ &  0.93/14         &  chi2 \\
	3C 196      & 15001         & $1.67_{-0.38}^{+0.40}$ & $1.75_{-0.39}^{+0.41}$ &  0.89/68         &  cstat           & $1.22_{-0.62}^{+0.66}$ &  1.02/24         &  cstat \\
	3C 204      & 9248          & $1.95_{-0.14}^{+0.14}$ & $2.00_{-0.14}^{+0.14}$ &  0.98/154        &  cstat           & $1.78_{-0.16}^{+0.19}$ &  0.99/75         &  cstat \\
	3C 205      & 9249          & $1.66_{-0.09}^{+0.09}$ & $1.68_{-0.09}^{+0.09}$ &  0.84/53         &  chi2            & $1.60_{-0.20}^{+0.21}$ &  1.15/14         &  chi2 \\
	3C 207      & 2130          & $1.59_{-0.23}^{+0.08}$ & $1.56_{-0.21}^{+0.10}$ &  0.95/147        &  chi2            & $1.37_{-0.09}^{+0.09}$ &  1.00/51         &  chi2 \\
	3C 208      & 9250          & $1.63_{-0.14}^{+0.18}$ & $1.63_{-0.08}^{+0.13}$ &  0.91/15         &  chi2            & $1.54_{-0.14}^{+0.24}$ &  0.63/69         &  cstat \\
	3C 212      & 434           & $1.68_{-0.04}^{+0.04}$ & $1.67_{-0.04}^{+0.04}$ &  0.74/159        &  chi2            & $1.57_{-0.08}^{+0.08}$ &  1.22/60         &  chi2 \\
	3C 215      & 3054          & $1.88_{-0.08}^{+0.09}$ & $1.87_{-0.08}^{+0.09}$ &  0.94/223        &  chi2            & $1.69_{-0.05}^{+0.06}$ &  0.92/124        &  chi2 \\
	3C 216      & 15002         & $1.59_{-0.13}^{+0.26}$ & $1.62_{-0.15}^{+0.25}$ &  1.14/12        &  chi2           & $1.41_{-0.17}^{+0.27}$ &  0.93/63         &  cstat \\
	3C 245      & 2136          & $1.55_{-0.03}^{+0.06}$ & $1.55_{-0.03}^{+0.06}$ &  1.06/95         &  chi2            & $1.47_{-0.09}^{+0.11}$ &  0.74/33         &  chi2 \\
	3C 249.1    & 3986          & $1.92_{-0.04}^{+0.13}$ & $1.99_{-0.08}^{+0.07}$ &  1.09/188        &  chi2            & $1.60_{-0.05}^{+0.05}$ &  0.98/93         &  chi2 \\
	3C 254      & 2209          & $1.93_{-0.07}^{+0.10}$ & $1.95_{-0.08}^{+0.09}$ &  0.90/192        &  chi2            & $1.49_{-0.03}^{+0.06}$ &  1.05/101        &  chi2 \\
	3C 263      & 2126          & $1.83_{-0.05}^{+0.09}$ & $1.81_{-0.08}^{+0.08}$ &  1.02/197        &  chi2            & $1.58_{-0.07}^{+0.05}$ &  1.57/78         &  chi2 \\
	3C 268.4    & 9325          & $1.45_{-0.12}^{+0.17}$ & $1.53_{-0.14}^{+0.14}$ &  0.90/160        &  cstat           & $1.18_{-0.15}^{+0.22}$ &  0.95/70         &  cstat \\
	3C 270.1    & 13906         & $1.57_{-0.07}^{+0.07}$ & $1.58_{-0.07}^{+0.07}$ &  1.00/243        &  chi2            & $1.42_{-0.05}^{+0.05}$ &  0.91/118        &  chi2 \\
	3C 275.1    & 2096          & $1.79_{-0.10}^{+0.11}$ & $1.79_{-0.10}^{+0.11}$ &  1.01/166        &  chi2            & $1.56_{-0.07}^{+0.07}$ &  0.86/66         &  chi2 \\
	3C 287      & 3103          & $1.86_{-0.04}^{+0.04}$ & $1.84_{-0.04}^{+0.04}$ &  0.86/140        &  chi2            & $1.70_{-0.08}^{+0.08}$ &  0.98/60         &  chi2 \\
	3C 286      & 15006         & $2.12_{-0.26}^{+0.51}$ & $2.14_{-0.15}^{+0.23}$ &  0.67/86         &  cstat           & $2.35_{-0.38}^{+0.53}$ &  0.94/27         &  cstat \\
	3C 309.1    & 3105          & $1.46_{-0.01}^{+0.09}$ & $1.45_{-0.01}^{+0.10}$ &  0.77/187        &  chi2            & $1.47_{-0.04}^{+0.04}$ &  1.09/87         &  chi2 \\
	3C 325      & 4818          & $1.45_{-0.20}^{+0.21}$ & $1.43_{-0.22}^{+0.23}$ &  0.81/18         &  chi2            & $0.38_{-0.13}^{+0.19}$ &  1.09/89         &  cstat \\
	3C 334      & 2097          & $1.85_{-0.11}^{+0.18}$ & $1.84_{-0.10}^{+0.18}$ &  1.17/52         &  chi2            & $1.82_{-0.12}^{+0.12}$ &  1.55/17         &  chi2 \\
	3C 336      & 15008         & $1.82_{-0.18}^{+0.18}$ & $1.85_{-0.19}^{+0.18}$ &  0.94/124        &  cstat           & $1.87_{-0.20}^{+0.36}$ &  1.05/50         &  cstat \\
	3C 345      & 2143          & $1.69_{-0.02}^{+0.09}$ & $1.68_{-0.02}^{+0.09}$ &  1.08/190        &  chi2            & $1.50_{-0.02}^{+0.08}$ &  0.85/96         &  chi2 \\
	3C 351      & 435           & $1.66_{-0.03}^{+0.70}$ & $1.65_{-0.03}^{+0.75}$ &  1.10/112        &  chi2            & $1.80_{-0.12}^{+0.36}$ &  1.03/38         &  chi2 \\
	4C 16.49    & 9262          & $1.83_{-0.12}^{+0.20}$ & $1.86_{-0.14}^{+0.21}$ &  0.90/111        &  cstat           & $1.65_{-0.23}^{+0.31}$ &  1.06/49         &  cstat \\
	3C 380      & 3124          & $1.76_{-0.10}^{+0.12}$ & $1.67_{-0.10}^{+0.15}$ &  0.99/129        &  chi2            & $1.45_{-0.04}^{+0.09}$ &  0.97/57         &  chi2 \\
	3C 432      & 5624          & $1.76_{-0.10}^{+0.10}$ & $1.77_{-0.10}^{+0.10}$ &  1.14/39         &  chi2            & $1.56_{-0.21}^{+0.22}$ &  0.96/15         &  chi2 \\
	3C 454        & 21403         & $1.75_{-0.19}^{+0.20}$ & $1.78_{-0.20}^{+0.21}$ & 0.92/22       & chi2          & $1.61_{-0.26}^{+0.59}$ & 0.99/106      & cstat \\
	3C 454.3    & 4843          & $1.67_{-0.05}^{+0.06}$ & $1.65_{-0.06}^{+0.05}$ &  1.01/383        &  chi2            & $1.35_{-0.03}^{+0.03}$ &  0.98/225        &  chi2 \\
	3C 455      & 15014         & $1.65_{-0.21}^{+0.12}$ & $1.57_{-0.14}^{+0.19}$ &  0.78/108        &  cstat           & $1.43_{-0.22}^{+0.42}$ &  1.07/49         &  cstat \\
	\enddata
	\tablecomments{Column (1): 3CRR name; Column (2): Chandra observation ID; Column (3): Photon index of power-law component extracted from source-centered $2.5\arcsec$ circle; Column ($ 4-6 $): Photon index, reduced $\chi^2$/degree of freedom and statistical method for the power-law fit in $1.5\arcsec$ circle; Columns ($ 7-9 $): The power-law fit in $0.5-2.0\arcsec$ annulus.}
\end{deluxetable*}

\begin{figure}[h]
	\includegraphics[width=1.0\columnwidth]{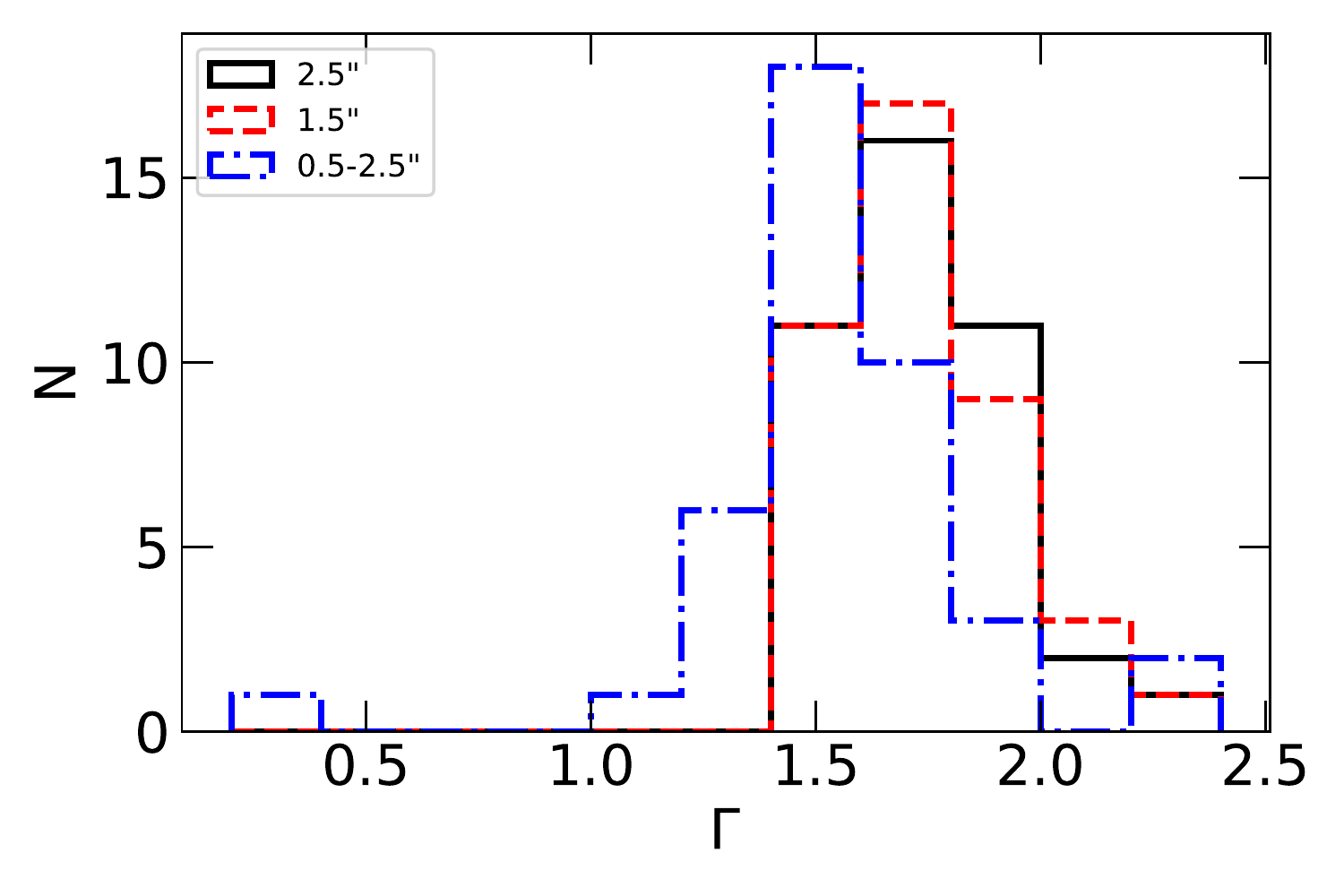}
	\caption{The distribution of X-ray photon index $ \Gamma $. The solid line is for $2.5\arcsec$ region, and the dashed line represents $1.5\arcsec$ region.
		As comparison, the $0.5-2.0\arcsec$ annulus is shown as dot-dashed line. \label{fig:gammahist} }
\end{figure}

\begin{figure}[h]
	\includegraphics[width=1.0\columnwidth]{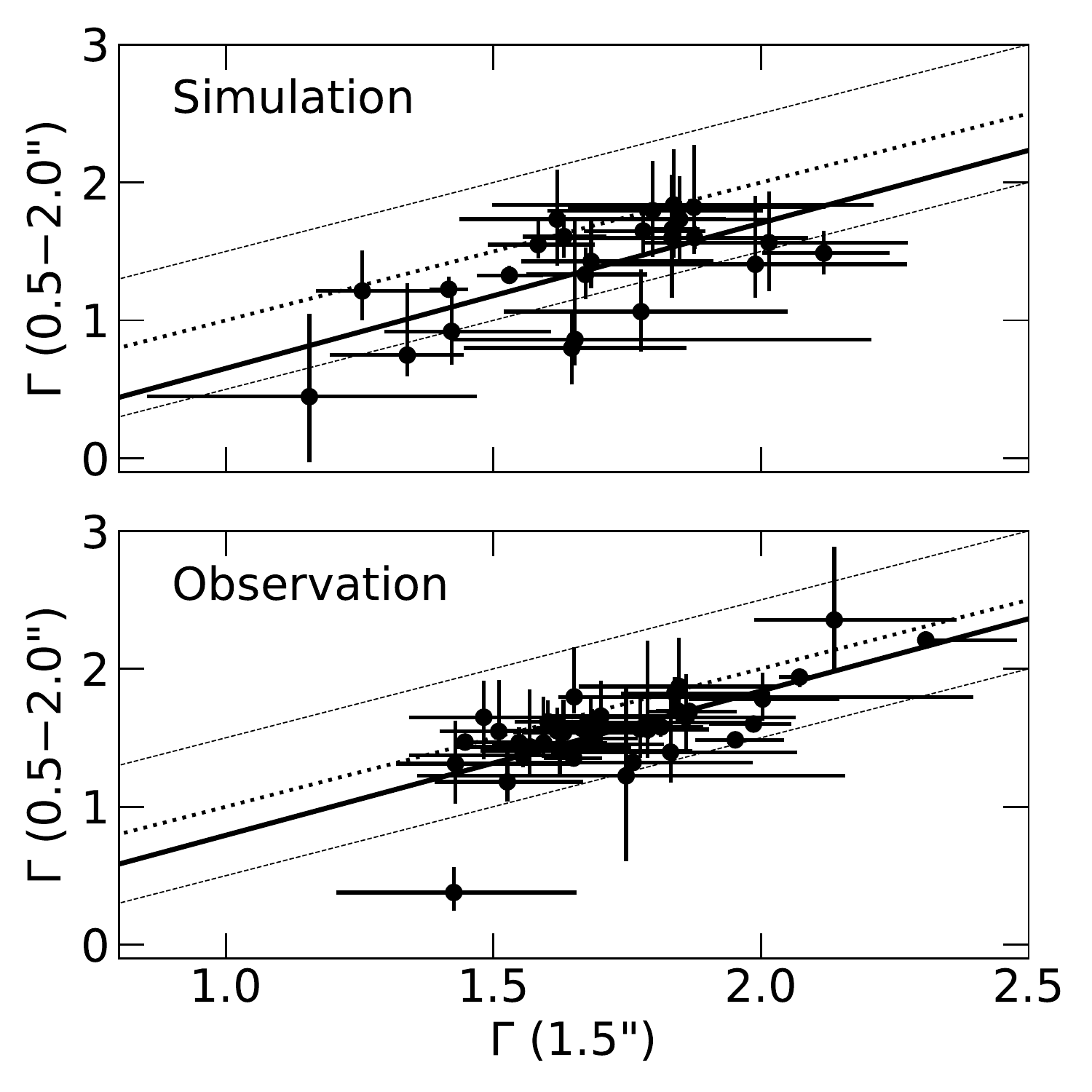}
	\caption{The relation of the X-ray photon index from $1.5\arcsec$ circle and from $0.5-2.0\arcsec$ annulus: ($upper$) - simulation; ($bottom$) - observation. The black dotted line is equivalent line, and its 0.5 offset is shown as dashed lines. The black solid lines show the linear fit. \label{fig:spec-gamma}}
\end{figure}

\end{document}